\documentclass[aps,pra,twocolumn,showpacs,superscriptaddress,10pt]{revtex4-1}

\pdfoutput=1

\usepackage[T1]{fontenc}
\usepackage[utf8]{inputenc}
\usepackage{amsmath}
\usepackage{amsfonts}
\usepackage{amssymb}
\usepackage{mathrsfs}
\usepackage{graphicx}
\usepackage{color}
\usepackage[usenames,dvipsnames]{xcolor}

\newcommand{\xx}{\mathbf{x}}

\newcommand{\rr}{\mathbf{r}}

\newcommand{\dd}{\mathrm{d}}
\newcommand{\ee}{\mathrm{e}}
\newcommand{\kk}{\mathbf{k}}

\newcommand{\vecpi}{\boldsymbol\Pi}
\newcommand{\vecphi}{\boldsymbol\phi}
\newcommand{\vecvarphi}{\boldsymbol\varphi}

\begin{document}
\title{{Aging and coarsening in isolated quantum systems after a quench: exact results for the quantum $O(N)$ model with $N \to \infty$}}
\author{Anna Maraga}
\affiliation{SISSA --- International School for Advanced Studies and INFN, via Bonomea 265, I-34136 Trieste, Italy}
\author{Alessio Chiocchetta}
\affiliation{SISSA --- International School for Advanced Studies and INFN, via Bonomea 265, I-34136 Trieste, Italy}
\author{Aditi Mitra}
\affiliation{Department of Physics, New York University, 4 Washington Place, New York, NY 10003, USA}
\author{Andrea Gambassi}
\affiliation{SISSA --- International School for Advanced Studies and INFN, via Bonomea 265, I-34136 Trieste, Italy}

\begin{abstract}
The non-equilibrium dynamics of an isolated quantum system after a sudden quench to a dynamical critical point is expected to be characterized by scaling and universal exponents due to the absence of time scales. We explore these features for a quench of the parameters of a Hamiltonian with $O(N)$ symmetry, starting from a ground state in the disordered phase. In the limit of infinite $N$, the exponents and scaling forms of the relevant two-time correlation functions can be calculated exactly.
Our analytical predictions are confirmed by the numerical solution of the corresponding equations. Moreover, we find that the same scaling functions, yet with different exponents, also describe the coarsening dynamics for quenches below the dynamical critical point.
\end{abstract}


\date{\today}
\maketitle

\section{Introduction}
\label{sec:introduction}
Understanding the non-equilibrium dynamics of isolated macroscopic quantum systems
is currently one of the most intriguing challenges of statistical physics~\cite{PolkovnikovRMP}. The interest in this problem has been recently revived by impressive advances in the physics of ultracold atoms~\cite{Lamacraft2012,Yukalov2011,Bloch2008,Greiner2002},
which paved the way to the experimental investigation of the real-time dynamics of almost isolated
quantum many-body systems~\cite{GreinerMandel02,Trotzky2008,Bakr2009,Bakr2010,Sherson2010,Endres2011,Cheneau2012,Fukuhara2013,Betz2011,Langen2013} with tunable properties such as interactions, dimensionality, internal symmetries, etc.
Consequently, what was an academic question until very recently, i.e., whether and how an isolated quantum system thermalizes after a \emph{quench} of the parameters of its Hamiltonian, became a subject of extensive investigation~\cite{Deutsch1991,Srednicki1994,Rigol2008}.
Remarkably, the study of the approach to thermal equilibrium disclosed a number of novel features such as the existence of \emph{prethermal} quasi-stationary states, predicted in Ref.~\cite{Berges2004}, and subsequently experimentally observed in Refs.~\cite{Kitagawa2011,Gring2012,Langen2013}.
It was then argued~\cite{Kollar2011,Moeckel2008,Moeckel2009,Moeckel2010,Marino2012,Marcuzzi2013,Mitra2013,VanDenWorm2013,Bertini2015} that this prethermalization occurs due to the system retaining memory of the integrable part of the Hamiltonian, which, in the absence of non-integrable terms, would lead the system towards a non-thermal
state. The question of whether these prethermal states can be represented by a generalized Gibbs ensemble is also a topic of current, active research~\cite{Rigol2007,Iucci2009,Jaynes1957,Barthel2008,Goldstein2014,Pozsgay2014,Mierzejewski2014,Wouters2014}.

It has also been argued that a quantum system can undergo a \emph{dynamical phase transition} (DPT) ~\cite{Sciolla2010,Gambassi2011,Sciolla2011,Sciolla2013,Sondhi2013,Smacchia2014,Eckstein2009,Schiro2010} in the prethermal regime, with some of the features of an equilibrium quantum phase transition, such as the appearance of diverging length- and time-scales. In particular, such a DPT was shown~\cite{Smacchia2014,Chiocchetta2015} to be related, in some respects, to an equilibrium one with an effective temperature proportional to the amount of energy injected by the quench into the system.
On this basis, one expects some degree of \emph{universality} to emerge close to a DPT, with spatial and temporal correlations characterized by critical exponents which do not depend on the microscopic details of the system. In particular, the lack of time scales combined with the breaking of translational symmetry in time due to the sudden quench is expected to cause an algebraic relaxation of the system to its stationary state, a phenomenon which has been referred to as \emph{quantum aging}, in analogy with the slow dynamics observed in glasses~\cite{Cugliandolo1997}.
While aging is known to occur in classical critical systems~\cite{Janssen1988,Gambassi2005}, only recently the same issue has been investigated after quenches in open~\cite{Gagel2014,Gagel2015,Buchold2014} or isolated~\cite{Chiocchetta2015} quantum systems. However, while in the former case the critical point responsible for aging is the thermal one dictated by the presence of a thermal bath, in the latter the critical point has an intrinsic non-equilibrium nature.

The consequences of this difference has been recently explored in Ref.~\cite{Chiocchetta2015}, where a system with vector order parameter $\vecphi$ with $N$ components and a Hamiltonian with $O(N)$ symmetry was studied by means of a perturbative renormalization-group approach. Aging was shown to occur in an early stage of the evolution, when the system exhibits prethermalization, rather than in the eventual thermalized state. This prethermal state was recognized to be similar to the non-equilibrium stationary state approached by the same model in the limit $N \to \infty$, 
in which it becomes integrable~\cite{Sondhi2013,Sciolla2013, Smacchia2014}. This fact suggests that the $O(N)$-symmetric Hamiltonian for $N \to \infty$ is the integrable Hamiltonian responsible for the prethermalization 
occurring for finite $N$.

In the present work we report a complementary non-perturbative analysis of the problem analyzed perturbatively in Ref.~\cite{Chiocchetta2015}, by considering the post-quench dynamics in the limit $N\rightarrow \infty$, when the model is numerically and analytically exactly solvable.

The structure of the paper is as follows: in Sec.~\ref{sec:model} we introduce the model and the protocol of the quench, and derive the relevant self-consistent evolution equations which are exact in the limit $N\rightarrow \infty$.
In Sec.~\ref{sec:scaling} we discuss their analytic solution in the absence of time- and length-scales in the system, i.e., after a quench at or below the dynamical critical point. In particular, we provide analytical predictions for the relevant scaling exponents and scaling functions for generic spatial dimensionality $d$ of the model.
In Sec.~\ref{sec:numerical} we present the results of the numerical solution of the evolution equations for a quench to the critical point (Sec.~\ref{sec:critical}) and below it (Sec.~\ref{sec:coarsening}), finding excellent agreement with the 
analytic predictions; for comparison we briefly discuss the latter case in the context of coarsening (Sec.~\ref{sec:coarsening-exp}). Finally, in Sec.~\ref{sec:conclusions} we summarize our findings and briefly discuss their possible implications for recent experimental proposals.

\section{Model and quench protocol}
\label{sec:model}
We consider a system described by the following $O(N)$-symmetric Hamiltonian in $d$ spatial dimensions
\begin{equation}
\label{eq:Hamiltonian}
H(r,u) = \int\dd^d x\, \left[\frac{1}{2}\vecpi^2 + \frac{1}{2}(\nabla\vecphi)^2 + \frac{r}{2}\vecphi^2 + \frac{u}{4! N}(\vecphi^2)^2 \right],
\end{equation}
where $\vecphi = (\phi_1,\dots,\phi_N)$ is a bosonic field with $N$ components, while $\vecpi$ is the canonically conjugated momentum with $[\phi_a(\xx),\Pi_b(\xx')] = i \delta^{(d)}(\xx-\xx')\delta_{ab}$. {Note that only the scalar products $\vecphi^2 = \vecphi\cdot\vecphi= \sum_{i=1}^N \phi_i^2$ and $\vecpi^2$ enter Eq. \eqref{eq:Hamiltonian}, as a consequence of the symmetry.}
The coupling $u>0$ controls the strength of the anharmonic interaction.
At time $t<0$ the system is prepared in the disordered ground state $|\psi_0\rangle$ of the pre-quench non-interacting Hamiltonian $H(\Omega_0^2,0)$, while the system is evolved with the post-quench Hamiltonian $H(r,u)$ for $t \ge 0$.
This protocol was studied in Ref.~\cite{Chiocchetta2015}, while in
Refs.~\cite{Gambassi2011,Sondhi2013,Sciolla2013,Smacchia2014} $u$ had the same non-zero value in the pre- and post-quench Hamiltonian. For the latter, it was shown that the system undergoes a \emph{dynamical phase transition}  (DPT) at a certain $r=r_c$, below which coarsening occurs.
As argued in Ref.~\cite{Chiocchetta2015},
a non-vanishing $u$ in the present initial Hamiltonian would only renormalize the value $\Omega_0^2$ and therefore we expect the model discussed here to undergo a DPT such as the one studied in
Refs.~\cite{Gambassi2011,Sondhi2013,Sciolla2013,Smacchia2014}.
The upper critical dimension of the DPT is $d_c = 4$, as in the corresponding equilibrium theory at finite temperature~\cite{Smacchia2014,Chiocchetta2015}. 
Accordingly, the critical, collective properties for $d>4$
are completely described by the corresponding non-interacting theory, as discussed in Sec.~\ref{sec:scaling}.
For $d$ smaller than the lower critical dimensionality $d_l = 2$~\cite{Smacchia2014}, instead,
this DPT is not expected to occur. However, since a counterpart of the Mermin-Wagner theorem is not known 
out of equilibrium, the occurrence of a different DPT cannot be ruled out for $d<d_l$.

In the following we will exploit the fact that, if the $O(N)$ symmetry in the initial state is not broken, the average $\langle \phi_a\phi_b \rangle$ (where $\langle \dots\rangle \equiv \langle \psi_0| \dots|\psi_0\rangle$) vanishes unless $a = b$, with its non-vanishing value independent of $a$, and equal to the fluctuation $\langle \phi^2 \rangle$ of a generic component $\phi$ of the field.
In the limit $N\to\infty$, the model can be solved by taking into account that, at the leading order, the quartic interaction in
Eq.~\eqref{eq:Hamiltonian}  decouples as~\cite{Sciolla2013,Sondhi2013,Smacchia2014,Sotiriadis2010}
\begin{equation}
(\vecphi^2)^2 \to 2(N+2)\langle \phi^2\rangle \vecphi^2 - N(N+2)\langle \phi^2\rangle^2;
\end{equation}
as shown in Ref.~\cite{Berges2007}, this decoupling corresponds to the Hartree-Fock approximation, which becomes exact for $N\to \infty$.
Once inserted into Eq.~\eqref{eq:Hamiltonian}, the dynamics of the various components of the fields decouples and each of them is ruled (up to an inconsequential additive constant) by the effective time-dependent quadratic Hamiltonian
\begin{equation}
\label{eq:effective-Hamiltonian}
H_\text{eff}(t) = \frac{1}{2} \int\dd^d x\, \left[\Pi^2 + (\nabla\phi)^2 + r_\text{eff}(t)\phi^2 \right],
\end{equation}
where $r_\text{eff}(t)$ is determined by the condition
\begin{equation}
\label{eq:def-reff}
r_\text{eff}(t) = r + \frac{u}{6}\langle \phi^2(\xx,t) \rangle.
\end{equation}
In a field-theoretical language,  $r_\text{eff}(t)$ plays the role of a renormalized square effective mass of the field $\phi$.
Due to the quadratic nature of $H_\text{eff}$, it is convenient to decompose the field $\phi$ into its
Fourier components  $\phi_\kk$, according to $\phi(\xx,t) = \int \dd^d k \, \phi_\kk(t)\ee^{i\kk\cdot\xx}/(2\pi)^d$, with an analogous decomposition for $\Pi$.
 Each of these components can be written in terms of the annihilation and creation operators~\cite{Sondhi2013,Smacchia2014} $a_\kk$ and $a^\dagger_\kk$, respectively, diagonalizing the initial
Hamiltonian:
\begin{equation}
\label{eq:f-definition}
\phi_\kk(t) = f_\kk(t)a_{\kk} + f^*_\kk(t)a^\dagger_{-\kk},
\end{equation}
where $f_\kk(t) = f_{-\kk}(t)$ is a complex function depending on both time $t$ and momentum $\kk$. Note that the canonical commutation relations between $\phi$ and $\Pi$ imply~\cite{Sondhi2013}
\begin{equation}
\label{eq:equation-canonical}
2 \,\text{Im}[f_\kk(t)\dot{f}_\kk^*(t)] = 1.
\end{equation}
The Heisenberg equation of motion for $\phi_\kk$ derived from the Hamiltonian \eqref{eq:effective-Hamiltonian} yields the following evolution equation for $f_\kk(t)$:
\begin{equation}
\label{eq:f-equation}
\ddot f_\kk  + [k^2 + r_\text{eff}(t)]f_\kk = 0,
\end{equation}
with $k= |\kk|$. This equation is supplemented by the initial conditions
\begin{equation}
\label{eq:in-cond}
f_\kk(0) = 1/\sqrt{2\omega_{0k}} \quad \mbox{and} \quad \dot{f}_\kk(0) = -i\sqrt{\omega_{0k}/2},
\end{equation}
where $\omega^2_{0k} = k^2 + \Omega_0^2$, which can be obtained by diagonalizing the quadratic pre-quench Hamiltonian and by imposing the continuity~\cite{Calabrese2006,Calabrese2007} of $\phi_\kk(t)$ at $t=0$.

Since the Hamiltonian is quadratic, all the information on the dynamics is encoded in its two-time functions, such as
the retarded ($G_R$) and Keldysh ($G_K$) Green's functions, which are defined by~\cite{Kamenevbook2011}
\begin{align}
\delta_{\kk,-\kk'} iG_R(k,t,t') & =\theta(t-t')\langle [\phi_\kk(t),\phi_{\kk'}(t')]\rangle ,   \label{eq:GR-def}\\
\delta_{\kk,-\kk'} iG_K(k,t,t') & = \langle \{\phi_\kk(t),\phi_{\kk'}(t')\}\rangle, \label{eq:GK-def}
\end{align}
where $\delta_{\kk,\kk'} \equiv (2\pi)^d\delta^{(d)}(\kk - \kk')$, while $\theta(t<0)=0$ and $\theta(t\ge 0)=1$.
Accordingly, $r_\text{eff}$ in Eq.~\eqref{eq:def-reff} can be expressed in terms of $G_K$ as
\begin{equation}
\label{eq:self-con1}
r_\text{eff}(t) = r + \frac{u}{12} \int \frac{\dd^d k}{(2\pi)^d} \, i G_K(k,t,t) h(k/\Lambda),
\end{equation}
%
%
where the function 
\begin{equation}
h(x) = 
\begin{cases}
0 & \mbox{for} \quad x \gg 1,\\
1 &\mbox{for} \quad x \ll 1,
\end{cases}
\label{eq:gen-h}
\end{equation}
implements a large-$\kk$ cut-off at the scale $\Lambda$
in order to make the theory well-defined at short distances, while it does not affect it for $k\ll \Lambda$.
By using Eq.~\eqref{eq:f-definition} in Eqs.~\eqref{eq:GK-def} and \eqref{eq:GR-def}, we find that $G_{K,R}$
can be written in terms of the function $f_\kk$ as:
\begin{align}
iG_K(k,t,t') & = 2 \text{Re} \left[ f_\kk(t)f^*_\kk(t') \right], \label{eq:GK-f}\\
G_R(k,t,t') & =  2 \theta(t-t')\text{Im}\left[ f_\kk(t)f^*_\kk(t')\right]. \label{eq:GR-f}
\end{align}
The dynamics of the system can be determined by solving the set of self-consistency equations given by Eqs.~\eqref{eq:f-equation}, \eqref{eq:self-con1} and \eqref{eq:GK-f}. Generically, these equations do not admit an analytic solution and therefore one has to resort to numerical integration. Nevertheless, in the following section, we show that some
quantities can be analytically calculated when the post-quench Hamiltonian is close to the dynamical critical point.

\section{Dynamical phase transition and scaling equations}
\label{sec:scaling}
In Refs.~\cite{Sondhi2013,Sciolla2013, Smacchia2014} it was shown that, after the quench, the system approaches a stationary state, in which the effective Hamiltonian becomes time-independent. Such a stationary state was then argued to be non-thermal as a consequence of the integrability of the model.
In particular, in Ref.~\cite{Smacchia2014}, it was demonstrated that when $r$ in Eq.~\eqref{eq:Hamiltonian} is tuned to a critical value $r_c$, the long-time limit $r^*$ of the corresponding effective parameter $r_\text{eff}(t)$ in Eq.~\eqref{eq:effective-Hamiltonian} vanishes and therefore the fluctuations of the order parameter become critical, signalling the occurrence of a dynamical phase transition. More precisely, as a consequence of the divergence of the spatial correlation length $\xi \equiv (r^*)^{-1/2}$, the correlation functions at long times acquire scaling forms characterized by universal critical exponents.

Similarly to the case of classical systems after a quench of the temperature~\cite{Gambassi2005},
the correlation functions exhibit dynamical scaling forms not only in the steady state, but also while approaching it~\cite{Chiocchetta2015}:
in particular, relying on dimensional analysis and on the lack of additional time- and length-scales at $r=r_c$, one expects the effective value $r_\text{eff}(t)$ to scale as:
\begin{equation}
\label{eq:ansatz0}
r_\text{eff}(t) = \frac{a}{t^2} \sigma(\Lambda t),
\end{equation}
where the function $\sigma$ is normalized by requiring $\sigma(\infty) = 1$, such that  $r_\text{eff}(t)$ vanishes at long times as
\begin{equation}
\label{eq:ansatz1}
r_\text{eff}(t) = \frac{a}{t^2} \quad \mbox{for}\quad \Lambda t \gg 1,
\end{equation}
while $a$ is a dimensionless quantity.
The non-universal  corrections introduced by $\sigma(\Lambda t)-1$ to this long-time limit are negligible for $\Lambda t \gg 1$.
On the contrary, for $\Lambda t \lesssim 1$, they become dominant and non-universal behavior is displayed.
Accordingly, one can identify a \emph{microscopic time}~\cite{Chiocchetta2015} $t_\Lambda \simeq \Lambda^{-1}$ which separates these two regimes: for $0 \leq t \lesssim t_\Lambda$ the dynamics is dominated by non-universal microscopic details; for $t\gtrsim t_\Lambda$, instead, the dynamics becomes universal.
This discussion assumes that the function $\sigma(\tau)$ has a well-defined limit as $\tau \to \infty$, which might not be the case in the presence of oscillatory terms. In fact, as shown in the numerical analysis presented in Sec.~\ref{sec:numerical}, the non-universal function $\sigma$ depends on how the cut-off $\Lambda$ is implemented in the model, i.e., on the choice of the function $h(x)$ in Eq. \eqref{eq:self-con1}. In particular, the choice of a sharp cut-off turns out to make $\sigma(\tau)$ oscillate, masking
the universal long-time behavior $r_\text{eff}(t)\sim t^{-2}$.

As a consequence of the universal form of Eq. \eqref{eq:ansatz1} for $t\gtrsim t_\Lambda$, the correlation functions are expected to exhibit scaling properties. In order to show this, it is convenient to rescale time and write the function $f_\kk(t)$ as $f_\kk(t) = g_\kk(kt)$. Inserting Eq.~\eqref{eq:ansatz1} into Eq. \eqref{eq:f-equation}, one finds the equation for $g_\kk(x)$:
\begin{equation}
\label{eq:g-equation}
g_\kk''(x) + \left(1 + \frac{a}{x^2} \right)g_\kk(x) = 0,
\end{equation}
valid for $x \equiv k t \gtrsim k t_\Lambda$, whose solution is:
\begin{equation}
\label{eq:g-solution}
g_\kk(x) = \sqrt{x}[A_\kk  J_\alpha(x) +B_\kk J_{-\alpha}(x)],
\end{equation}
where $J_\alpha(x)$ is the Bessel function of the first kind and
\begin{equation}
\label{eq:def-alpha}
\alpha = \sqrt{\frac{1}{4}-a}.
\end{equation}
Below we show that it is consistent to assume $a<1/4$ and therefore $\alpha$ to be real.  The constants $A_\kk$ and $B_\kk$ in Eq.~\eqref{eq:g-solution} are fixed by the initial condition of the evolution, as discussed below.
For later reference, we recall that
\begin{equation}
\label{eq:exp-J}
J_\alpha(x) \simeq \begin{cases}
(x/2)^\alpha/\Gamma(1+\alpha), &  x \ll 1,\\
\cos(x - \alpha \pi/2 -\pi/4) \sqrt{2/(\pi x)}, & x \gg 1,
\end{cases}
\end{equation}
where $\Gamma(x)$ is the Euler gamma function~\cite{Abramowitzbook}.
Note that Eq.~\eqref{eq:g-solution} encodes the complete dependence of $f_\kk(t)$ on time $t$ for $t \gtrsim t_\Lambda$, whereas its dependence on the wave vector $\kk$ is encoded in the yet unknown functions $A_\kk$ and $B_\kk$.
By using the following identity for the Wronskian of Bessel functions~\cite{Abramowitzbook}
\begin{equation}
J_\alpha(x)J'_{-\alpha}(x) - J_{-\alpha}(x)J_\alpha'(x) = -\frac{2\sin(\alpha\pi)}{\pi x},
\end{equation}
one can show that Eq.~\eqref{eq:equation-canonical} requires the coefficients $A_\kk$, $B_\kk$ to satisfy the relation:
\begin{equation}
\label{eq:AB-canonical}
\text{Im}[A_\kk B^*_\kk] = -\frac{\pi}{4\sin(\alpha\pi)} \frac{1}{k}.
\end{equation}
This relation is not sufficient in order to determine completely $A_\kk$ and $B_\kk$ unless the full functional form of $r_\text{eff}(t)$ is taken into account, including its non-universal behavior for $t\lesssim t_\Lambda$; this would allow us to fix $A_\kk$ and $B_\kk$ on the basis of the initial conditions for the evolution, which at present cannot be reached from Eq.~\eqref{eq:g-solution}, it being valid only for $t\gtrsim t_\Lambda$.
However, for a deep quench  --- such as that one investigated in Ref.~\cite{Chiocchetta2015}  ---
with $\Omega_0 \gg \Lambda$, the initial conditions \eqref{eq:in-cond}  for the evolution of $f_\kk$ become essentially independent of $\kk$ and read:
\begin{equation}
f_\kk(0) \simeq 1/\sqrt{2\Omega_0}, \quad \dot{f}_\kk(0) \simeq -i\sqrt{\Omega_0/2}.
\label{eq:in-cond-dq}
\end{equation}
At time $t \simeq t_\Lambda \simeq \Lambda^{-1}$, $f_\kk(t_\Lambda)$ can be calculated from a series expansion $f_\kk(t_\Lambda) = f_\kk(0) + t_\Lambda \dot{f}_\kk(0) + t_\Lambda^2 \ddot f_\kk(0) + {\mathcal O}(t_\Lambda^3)$ and by using Eqs.~\eqref{eq:f-equation} and \eqref{eq:in-cond-dq} one may readily conclude that its dependence on $k$ comes about via $(k/\Lambda)^2$ and $k/(\Omega_0\Lambda)$; accordingly, at the leading order, it can be neglected for $k \ll \Lambda$. On the other hand,
$f_\kk$ at $t_\Lambda$ can be evaluated from Eq.~\eqref{eq:g-solution} and, since $kt_\Lambda \simeq k/\Lambda \ll 1$, we can use the asymptotic form of the Bessel functions for small arguments, finding:
\begin{equation}
f_\kk(t_\Lambda) \propto A_\kk (kt_\Lambda)^{1/2+\alpha} + B_\kk (kt_\Lambda)^{1/2 -\alpha}.
\end{equation}
In order to have $f_\kk(t_\Lambda)$ independent of $\kk$ at the leading order, it is then necessary that:
\begin{equation}
\label{eq:AB-scaling}
A_\kk \simeq A (k/\Lambda)^{-1/2-\alpha}, \quad B_\kk \simeq B (k/\Lambda)^{-1/2+\alpha},
\end{equation}
where $A$ and $B$ are yet unknown complex numbers. While this scaling is expected to be true for $k \ll \Lambda$, non-universal corrections may appear for $k \simeq \Lambda$. Note that Eq.~\eqref{eq:AB-scaling} is consistent with \eqref{eq:AB-canonical}, provided that:
\begin{equation}
\text{Im}[A B^*] = -\frac{\pi \Lambda^{-1}}{4\sin(\alpha\pi)}.
\end{equation}
The numerical analysis discussed in Sec.~\ref{sec:numerical} actually shows that $A_\kk$ turns out to be purely imaginary, while $B_\kk$ real.
Combining Eqs.~\eqref{eq:AB-scaling} with \eqref{eq:g-solution}, \eqref{eq:GK-f}, and \eqref{eq:GR-f}, one finds a simple form for the retarded Green's function
\begin{equation}
\begin{split}
G_R(k,t,t') & = - \theta(t-t') \frac{\pi }{2\sin \alpha\pi}  (tt')^{1/2} \times  \\
& \quad \left[J_\alpha(kt)J_{-\alpha}(kt')  - J_{-\alpha}(kt)J_\alpha(kt')  \right],
\end{split}
\label{eq:GR-fin}
\end{equation}
while the one for $iG_K(k,t,t')$ is somewhat lengthy and thus we do not report it here explicitly. In order to determine $a$ and therefore $\alpha$ from the self-consistent condition in Eq.~\eqref{eq:self-con1}, it is actually sufficient to know $iG_K(k,t,t')$ for $t=t'$, which is given by
\begin{equation}
\begin{split}
\label{eq:GK-solution}
iG_K&(k,t,t)	= 2\Lambda t \left\{ |A|^2(k/\Lambda)^{-2\alpha}J^2_\alpha(kt) \right.\\
	&+ \left.|B|^2 (k/\Lambda)^{2\alpha} J^2_{-\alpha}(kt) + 2 \text{Re}[AB^*] J_\alpha(kt)J_{-\alpha}(kt)\right\}.
\end{split}
\end{equation}
Accordingly, while Eq.~\eqref{eq:AB-canonical} is sufficient in order to determine the complete form of $G_R$, the one of $G_K$ still contains unknown coefficients $A$ and $B$ which are eventually determined by the initial conditions; nevertheless, the  scaling properties of both of these functions are already apparent. In fact, their dynamics is characterized by two temporal regimes, which we refer to as \emph{short} ($kt \ll 1$) and \emph{long} ($kt \gg 1$) \emph{times}.
Stated differently, the temporal evolution of each mode $\kk$ has a typical time scale $\sim k^{-1}$ determined by the value of the momentum itself and the corresponding short-time regime extends to macroscopically long times for vanishing momenta.

Let us focus on $G_R$ in Eq.~\eqref{eq:GR-fin}: at short times $t'<t \ll k^{-1}$ it becomes independent of  $k$
\begin{equation}
\label{eq:GR-short-time}
G_R(k,t,t') \simeq -\frac{t}{2\alpha} \left (\frac{t'}{t} \right)^{1/2-\alpha} \left[ 1 - \left (\frac{t'}{t} \right)^{2\alpha}\right],
\end{equation}
where we used the asymptotic expansion in Eq.~\eqref{eq:exp-J}.
For well-separated times $t\gg t'$, the second terms in brackets is negligible and
$G_R(k,t\gg t',t')$ displays an algebraic dependence on the ratio $t'/t$.
At long times $k^{-1} \ll t', t$, instead, $G_R$ becomes time-translational invariant and by
keeping the leading order of the asymptotic expansion of the Bessel functions in Eq.~\eqref{eq:exp-J},
it reads
\begin{equation}
G_R(k,t,t') \simeq - \theta(t-t')\frac{\sin{k(t-t')}}{k},
\label{eq:GR-long-time}
\end{equation}
which is nothing but the $G_R$ of a critical Gaussian Hamiltonian after a deep quench~\cite{Chiocchetta2015}.
Similarly, at short times $t\ll k^{-1}$ and to  leading order in $\Lambda t$, $G_K(k,t,t)$ reads
(see Eqs.~\eqref{eq:GK-solution} and \eqref{eq:exp-J})
\begin{equation}
\label{eq:GK-short-time}
iG_K(k,t,t) \simeq \frac{2^{1-2\alpha}\,|A|^2}{\Gamma^2(1+\alpha)} (\Lambda t)^{1+2\alpha},
\end{equation}
which is independent of the momentum $k$, as it is the case for $G_R$ within the same temporal regime.
At long times $t \gg k^{-1}$, instead, one finds
\begin{equation}
\label{eq:GK-long-time}
iG_K(k,t,t) \simeq \frac{2|A|^2}{\pi} \left(\frac{\Lambda}{k}\right)^{1+2\alpha},
\end{equation}
to leading order in $k/\Lambda$, where the possible oscillating terms have been neglected, as they are supposed to average to zero when an integration over momenta is performed. The resulting expression turns out to be time-independent and, contrary to what happens with $G_R$, it \emph{does not} correspond to the $G_K$ of a critical Gaussian theory after a deep quench~\cite{Chiocchetta2015}, reported further below in Eq.~\eqref{eq:GK-Gauss-dq}.
This fact should be regarded as a consequence of the non-thermal nature of the stationary state which is eventually reached by the system and which retains memory of the initial state.
Since the effective Hamiltonian \eqref{eq:effective-Hamiltonian} is Gaussian, it is tempting to interpret the anomalous momentum dependence $\sim k^{-1-2\alpha}$ in Eq.~\eqref{eq:GK-long-time} in terms of a quench in a truly Gaussian theory. In the latter case, it was shown~\cite{Calabrese2006,Calabrese2007,Sotiriadis2009} that, for deep quenches,
\begin{equation}
iG_K(k,t,t) \simeq \frac{\Omega_0}{2k^2},
\label{eq:GK-Gauss-dq}
\end{equation}
which is similar to an equilibrium distribution with an effective temperature $T_\text{eff} \simeq \Omega_0$.  Accordingly, Eq.~\eqref{eq:GK-long-time} can be regarded as resulting from a quench of a Gaussian theory with a momentum-dependent initial ``temperature'' $\Omega_0(k) \sim k^{1-2\alpha}$.

Equations \eqref{eq:GK-short-time} and \eqref{eq:GK-long-time} show that the term proportional to $|A|^2$ in the expression \eqref{eq:GK-solution} of $G_K(k,t,t)$ is dominant at both long and short times:
accordingly, we can neglect the remaining ones and the Keldysh Green's function with two different times $t$, $t'$
acquires the scaling form
\begin{equation}
\label{eq:GK-scaling}
iG_K(k,t,t') = 2|A|^2 \left(\frac{\Lambda}{k}\right)^{1+2\alpha} \sqrt{k^2tt'} J_\alpha(kt)J_\alpha(kt'),
\end{equation}
which can be derived from Eqs.~\eqref{eq:g-solution}, \eqref{eq:AB-scaling}, and \eqref{eq:GK-f}.
This expression can now be used in order to derive from the self-consistency equation \eqref{eq:self-con1} the value of the constant $a$ and therefore of the exponent $\alpha$ (see Eq.~\eqref{eq:def-alpha}) which characterizes the scaling forms \eqref{eq:GR-fin} and \eqref{eq:GK-scaling}.
For $\Lambda t \gg 1$ one can approximate $r_\text{eff}(t)$ with Eq. \eqref{eq:ansatz1} and, by assuming a sharp cut-off

\begin{equation}
h(x) = \theta(1-x),
\label{eq:def-h-sharp}
\end{equation}
Eq. \eqref{eq:self-con1} becomes
\begin{equation}
\label{eq:self-con2}
\frac{a}{t^2} = r_c + \frac{a_d}{6} |A|^2\,u\Lambda^d R_{d,\alpha} (\Lambda t)
\end{equation}
at the critical value $r_c$ of $r$.
In this expression we introduced the function
\begin{equation}
R_{d,\alpha}(x) = x\int_0^1\dd y \, y^{d-1-2\alpha} J_\alpha^2(x y),
\end{equation}
while $a_d = \Omega_d/(2\pi)^d$ with $\Omega_d = 2\pi^{d/2}/\Gamma(d/2)$ the $d$-dimensional solid angle.
Expanding $R_{d,\alpha}(x)$ for large argument $x\gg 1$, one finds that
\begin{equation}
\label{eq:R-exp}
R_{d,\alpha}(x) = W_{d,\alpha}(x) + c^{(0)}_{d,\alpha} + \frac{c^{(1)}_{d,\alpha}}{ x^{d-1-2\alpha}} + \frac{c^{(2)}_{d,\alpha}}{x^2} + {\mathcal O}\left(\frac{1}{x^4}\right),
\end{equation}
where $W_{d,\alpha}(x)$ is a fast oscillating function which is a consequence of the sharp cut-off considered in the integral over momenta and $c^{(i)}_{d,\alpha}$ are certain coefficients, the relevant values of which are provided further below.
Once the expansion \eqref{eq:R-exp} is plugged into Eq.~\eqref{eq:self-con2}, the r.h.s.~of the latter can be expanded in decreasing powers of $\Lambda t$ which have to match the term on its
l.h.s., resulting in a set of conditions fixing the values of $r_c$, $\alpha$, and $u$. Notice that the oscillations contained in $W_{d,\alpha}(x)$ would be compensated by sub-leading terms in the l.h.s. of Eq.~\eqref{eq:self-con2} (which are not reported, see the discussion after Eq.~\eqref{eq:ansatz1}), thus confirming the non-universal nature of $W_{d,\alpha}(x)$.
The value of $r_c$ is determined such as to cancel the constant contribution $\propto  c^{(0)}_{d,\alpha}$ on the r.h.s., while $\alpha$ has to be fixed such that $c^{(1)}_{d,\alpha} =0$ in order to cancel the term  $\propto (\Lambda t)^{1+2\alpha-d}$ which cannot be matched by the l.h.s.\ of Eq.~\eqref{eq:self-con2}
\footnote{In passing we mention that, alternatively, one might require the exponent of this term to equal the one on the l.h.s.\ of Eq.~\eqref{eq:self-con2}, i.e., $1+2\alpha-d = -2$. Correspondingly, the divergence of $c_{d,\alpha}^{(2)} \propto a/[2\pi  (3 + 2\alpha-d)]$ cancels the one of $c^{(1)}_{d,\alpha}$ reported in Eq.~\eqref{eq:gammas}: the eventual contribution $\propto t^{-2}$ on the r.h.s.\ of Eq.~\eqref{eq:self-con2} is finite but negative and it would therefore require an unphysical negative value of the coupling constant $u$ in order to match the term on the l.h.s.\ of the same equation.}.
This procedure --- typically used for solving this kind of self-consistency equations~\cite{Bray1977,Janssen1988,ZinnJustinbook} --- can be regarded as a systematic way of canceling terms which result from corrections to scaling and, therefore, it allows a comparison with the results obtained within a renormalization group approach (with some exceptions, see Ref.~\cite{ZinnJustinbook}).
For our purposes, it is sufficient to focus on the condition
\begin{equation}
\label{eq:gammas}
c^{(1)}_{d,\alpha} = \frac{1}{2\sqrt{\pi}} \frac{\Gamma\left(\frac{1}{2}-\frac{d}{2} +\alpha\right)\Gamma\left(\frac{d}{2}\right)}{\Gamma\left(1-\frac{d}{2} +\alpha\right)\Gamma\left(1-\frac{d}{2}+2\alpha\right)} =0,
\end{equation}
which is solved by requiring the argument of one of the two $\Gamma$ functions in the denominator 
%
%
%
to equal a non-positive integer.
The two corresponding infinite sets of solutions $\mathcal{S'}$ and $\mathcal{S}$ for $\alpha$ are given by
\begin{equation}
{\mathcal S}' = \left\{ \frac{d-2}{2}-n\right\}_{n\geq 0}, \quad  {\mathcal S} = \left\{ \frac{d-2}{4}-\frac{n}{2}\right\}_{n\geq 0},
\end{equation}
with integer $n$.
The physically relevant solution can be selected by requiring
$a$ to match its Gaussian value $a=0$ (or, equivalently, $\alpha=1/2$, see Eq.~\eqref{eq:def-alpha}) at the upper critical dimensionality $d_c=4$ of the model~\cite{Smacchia2014,Chiocchetta2015}
%
Note that this Gaussian value of $a$  can be easily inferred by inspecting the scaling behavior of $G_{R,K}$ for the quench towards a non-interacting theory (compare, e.g., Eqs. \eqref{eq:GK-long-time} and \eqref{eq:GK-Gauss-dq}).
Accordingly, one finds a single possible solution from $\mathcal S$, i.e.,
\begin{equation}
\label{eq:a-analytic}
\alpha = \frac{d-2}{4} \qquad \mbox{and}\qquad a = \frac{d}{4}\left(1-\frac{d}{4}\right),
\end{equation}
with $a>0$ for all values of $d$ between $d_l=2$ and $d_c=4$, while for $d > 4$ the Gaussian theory applies and therefore
\begin{equation}
\label{eq:a-analytic-mf}
\alpha = \frac{1}{2} \qquad \mbox{and}\qquad a = 0.
\end{equation}
For $d \geq d_c$, $c_{d,\alpha}^{(2)}\propto a$ vanishes and the leading temporal dependence of the r.h.s.\ of Eq.~\eqref{eq:self-con2} at long times is $\propto c^{(1)}_{d,\alpha=1/2} (\Lambda t)^{-(d-2)}$ and therefore
\begin{equation}
\label{eq:reff-above-ucd}
r_\text{eff}(t)  \propto  (\Lambda t)^{-(d-2)},
\end{equation}
up to oscillating terms, instead of the behavior $\propto t^{-2}$ in Eq. \eqref{eq:ansatz1}.
Equations \eqref{eq:a-analytic} and \eqref{eq:a-analytic-mf} together with Eqs.~\eqref{eq:GR-fin} and \eqref{eq:GK-scaling} completely characterize the scaling behavior of $G_{K,R}$ after a deep quench to the critical point. These results can be compared with the predictions of Ref.~\cite{Chiocchetta2015}, formulated with a dimensional expansion around the upper critical dimensionality $d_c$ of the $O(N)$ model, which is expected to reduce to the present one for $N\to\infty$. {There, the time dependence of $G_{K,R}$ for $kt,kt' \ll 1$ and $t' \ll t$ was parametrized in terms of the exponent $\theta$ as}
\begin{equation}
{G_R(k,t,t') \propto - t(t'/t)^\theta \quad \text{and} \quad  G_K \propto (tt')^{1-\theta},}
\end{equation}
{from which it follows that $\theta$ is related to the exponent $\alpha$ introduced in Eqs. \eqref{eq:GR-fin} and \eqref{eq:GK-scaling} by $\theta = 1/2 -\alpha$. Moreover, it was found in Ref.~\cite{Chiocchetta2015} that $\theta = \epsilon/4 + {\mathcal O}(\epsilon^2)$ with $\epsilon \equiv 4-d$, in agreement with Eq.~\eqref{eq:a-analytic}, which yields $\theta = 1 -d/4$ for $d<4$.}

We note here that among the remaining solutions of Eq.~\eqref{eq:gammas} in $\mathcal S$ and ${\mathcal S}'$ which do not match the Gaussian value at $d=d_c$, only one turns out to be compatible with having a positive value of the coupling constant $u$ in Eq.~\eqref{eq:self-con2}.
This solution belongs to  ${\mathcal S}'$, and is given by
\begin{equation}
\alpha_\text{co} = \frac{d-2}{2} \qquad \mbox{with} \qquad a_\text{co} = \frac{(3-d)(d-1)}{4};
\label{eq:exp-co}
\end{equation}
remarkably, it turns out to be related to the coarsening  occurring after a quench to $r<r_c$, as we argue and demonstrate further below. 

A comparison between these results and those for the corresponding exactly solvable classical quench is discussed in Appendix~\ref{sec:comparison}.

\section{Numerical results}
\label{sec:numerical}

In order to test the quality of the analytical predictions of the previous section, we studied in detail the numerical solution of the evolution equations~\eqref{eq:f-equation} for $f_\kk(t)$, under the constraint provided by Eq.~\eqref{eq:self-con1}, with $G_K$ given by Eq.~\eqref{eq:GK-f}. The numerical integration of these equations has been performed using an algorithm based on the Bulirsch-Stoer method~\cite{NumericalRecipes}, while the integrals over momentum $k$ have been computed using the extended Simpson's rule~\cite{NumericalRecipes} with a mesh of $7.5\times 10^4$ points.

In Sec.~\ref{sec:critical} we consider the case of a quench to the critical point, comparing the numerical results for the relevant correlation functions with the analytical predictions derived in Sec.~\ref{sec:scaling}. In Sec.~\ref{sec:coarsening} we present, instead, results for a quench  below the critical point, and show some numerical evidence of the emergence of scaling properties during coarsening.

\subsection{Quench to the critical point}
\label{sec:critical}
In order to determine the critical value $r_c$ of the parameter $r$ in Eq.~\eqref{eq:Hamiltonian} one can conveniently use the ansatz proposed in Ref.~\cite{Smacchia2014,Sotiriadis2010}, i.e.,
\begin{equation}
r_c = - \frac{u}{4!} \int \frac{d^d k}{(2\pi)^d} \frac{2k^2 + \Omega_0^2}{k^2\sqrt{k^2+\Omega_0^2}} \,h(k/\Lambda),
\label{eq:rc-ansatz}
\end{equation}
which turns out to correctly predict $r_c$ also beyond the case of a deep quench. The rationale behind this ansatz relies on the assumption that $G_K(k,t,t)$ at $t \to \infty$ has approximately the same form as for the case of a quench to $r=0$ in 
the non-interacting case $u=0$~\cite{Sotiriadis2010}.  
Although the validity of this assumption for a quench to the critical point is questionable because, as shown in Sec.~\ref{sec:scaling}, $G_K$ differs significantly 
from the non-interacting case $u=0$ (see, e.g., Ref.~\cite{Chiocchetta2015}), Eq.~\eqref{eq:rc-ansatz} 
anyhow provides accurate 
predictions for the value of $r_c$.

The accuracy of the ansatz 
\eqref{eq:ansatz1} for the long-time behavior of $r_\text{eff}(t)$ can be tested by calculating $r_\text{eff}(t)$ according to Eqs.~\eqref{eq:self-con1} and \eqref{eq:GK-f}, based on the numerical solution of the evolution equation \eqref{eq:f-equation} for $f_\kk(t)$.
%
%
%

%
\begin{figure*}[h]
\centerline{
	\begin{tabular}{ccc}
	\includegraphics[width=5.48cm]{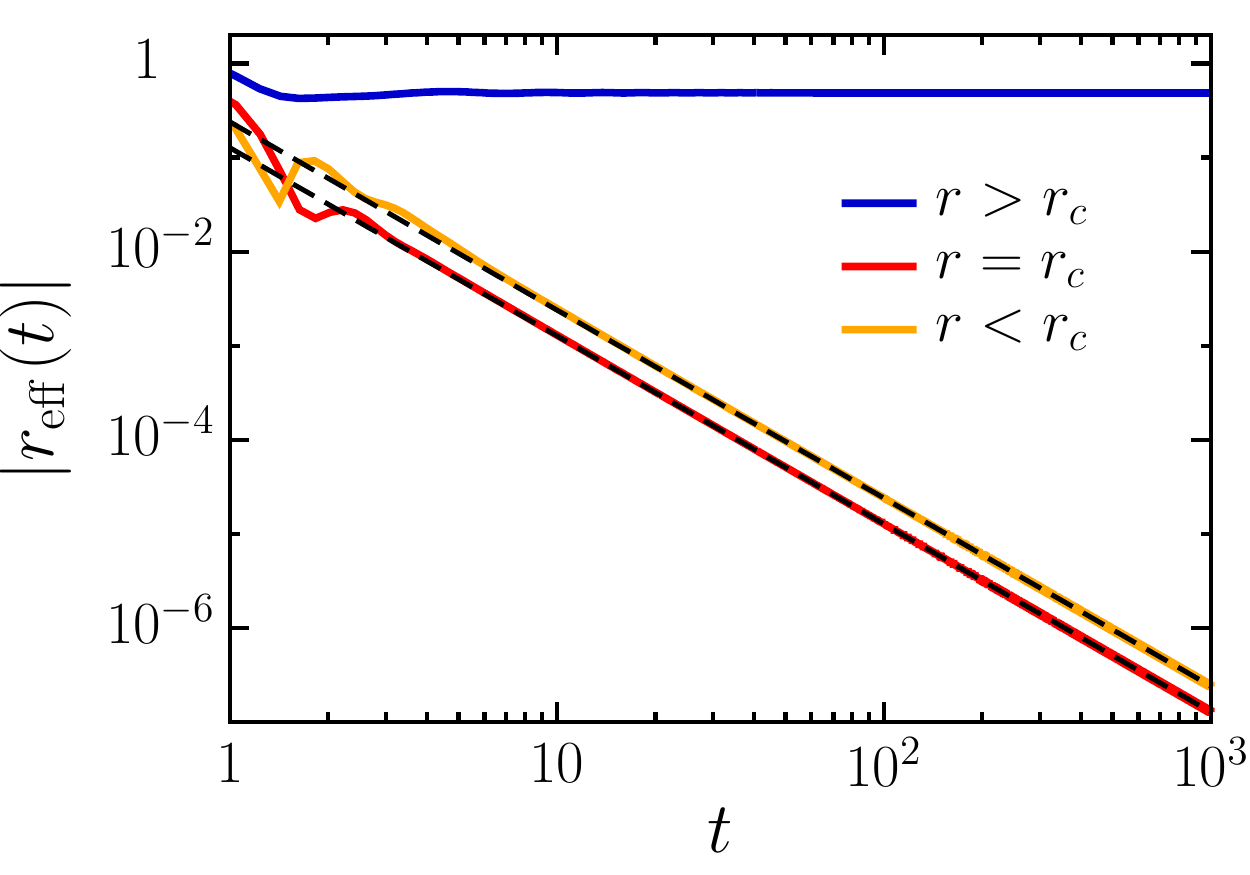} &
	\includegraphics[width=5.48cm]{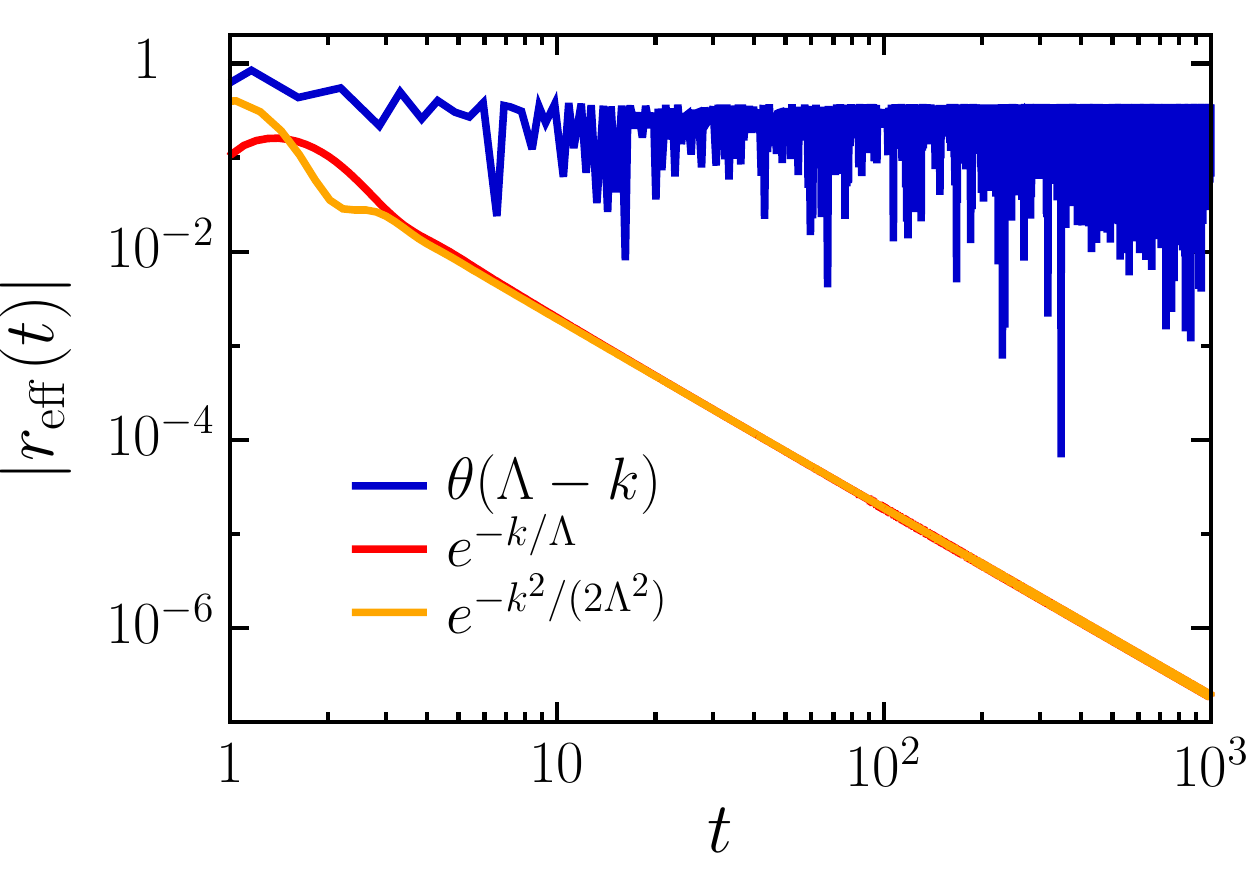} &
	\includegraphics[width=5.48cm]{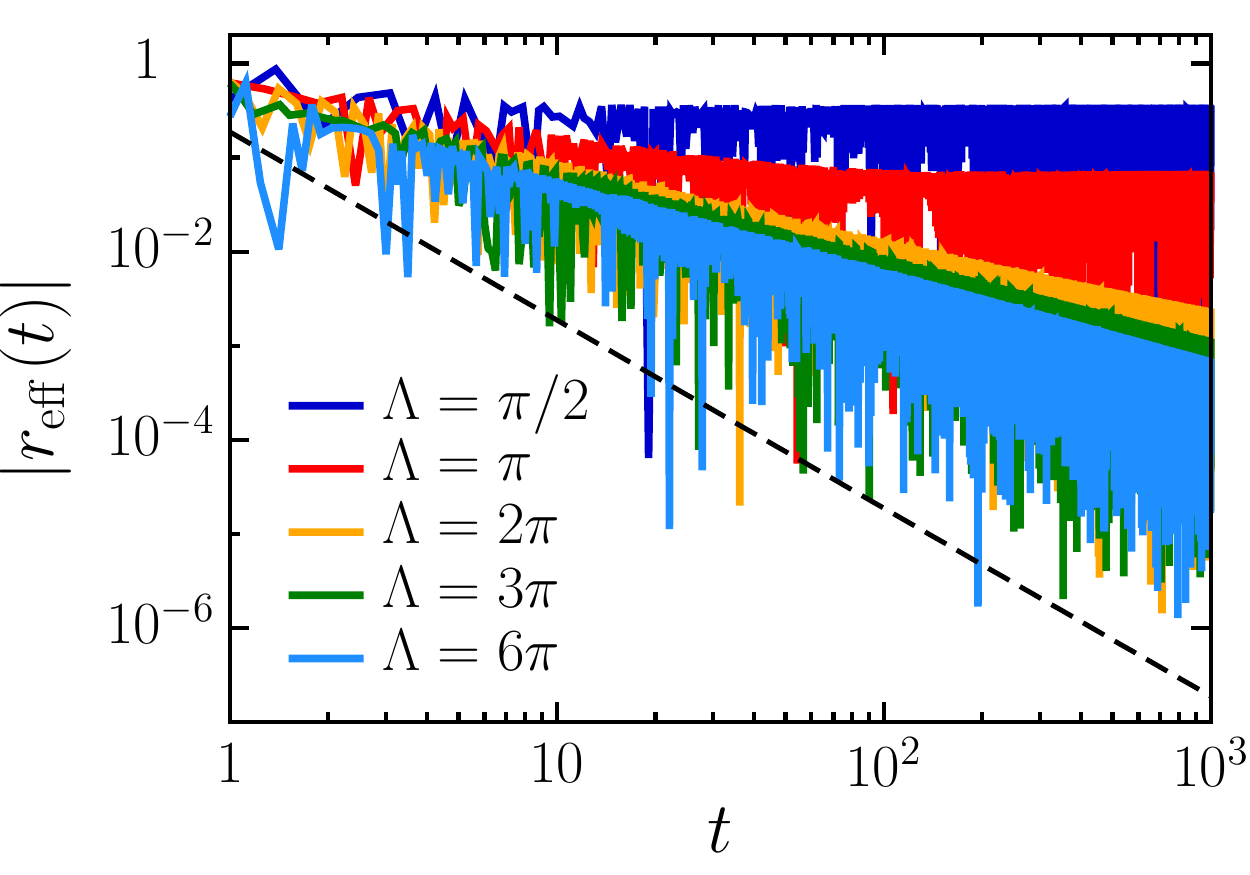} \\
	\includegraphics[width=5.48cm]{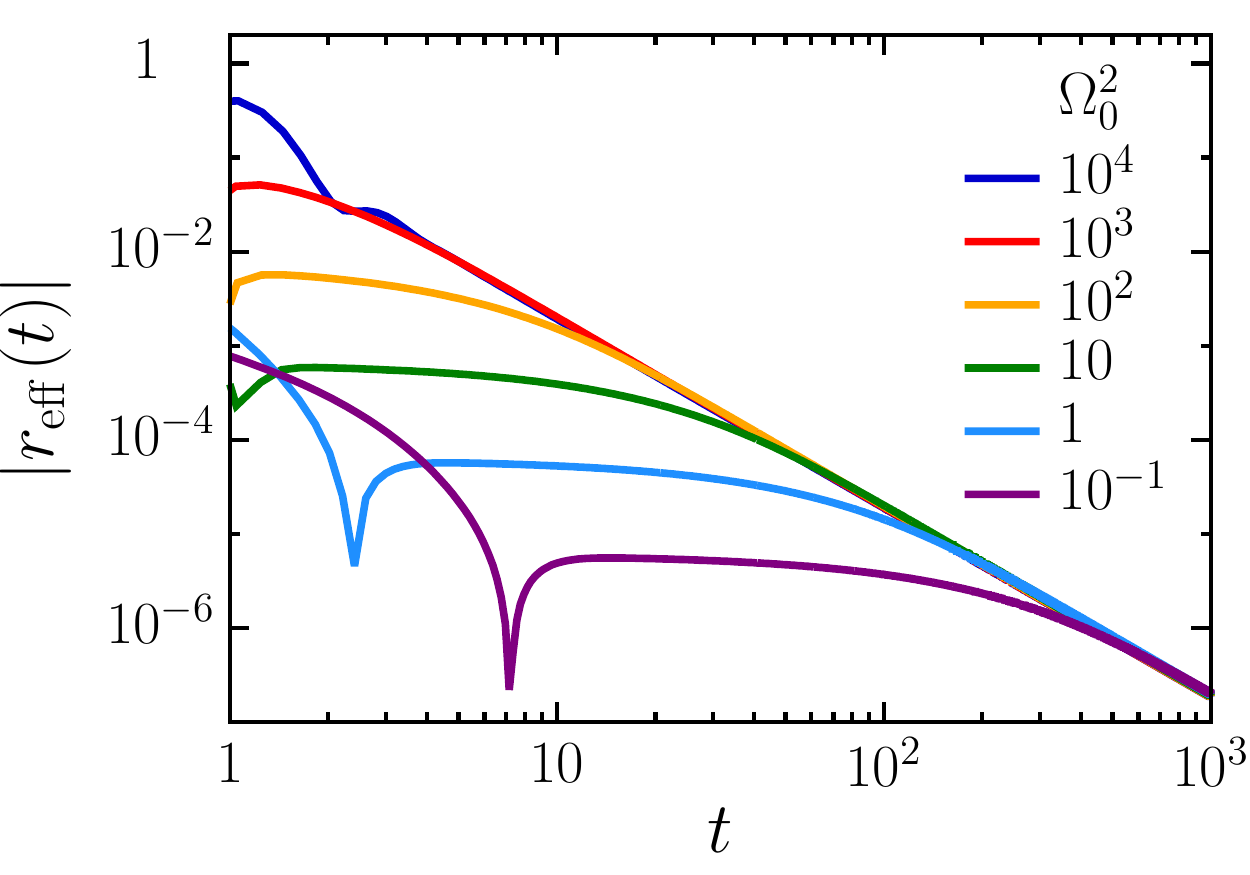} &
	\includegraphics[width=5.48cm]{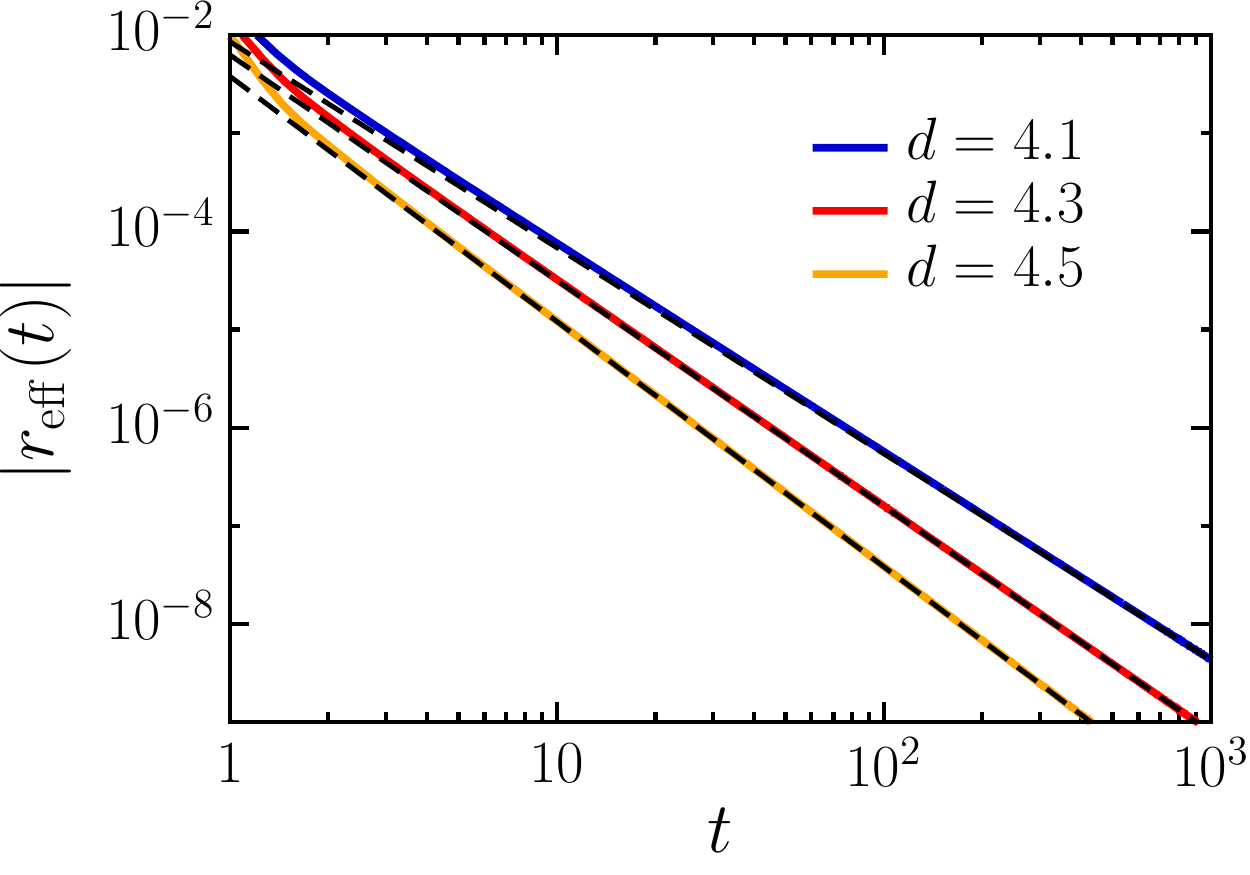} &
	\includegraphics[width=5.48cm]{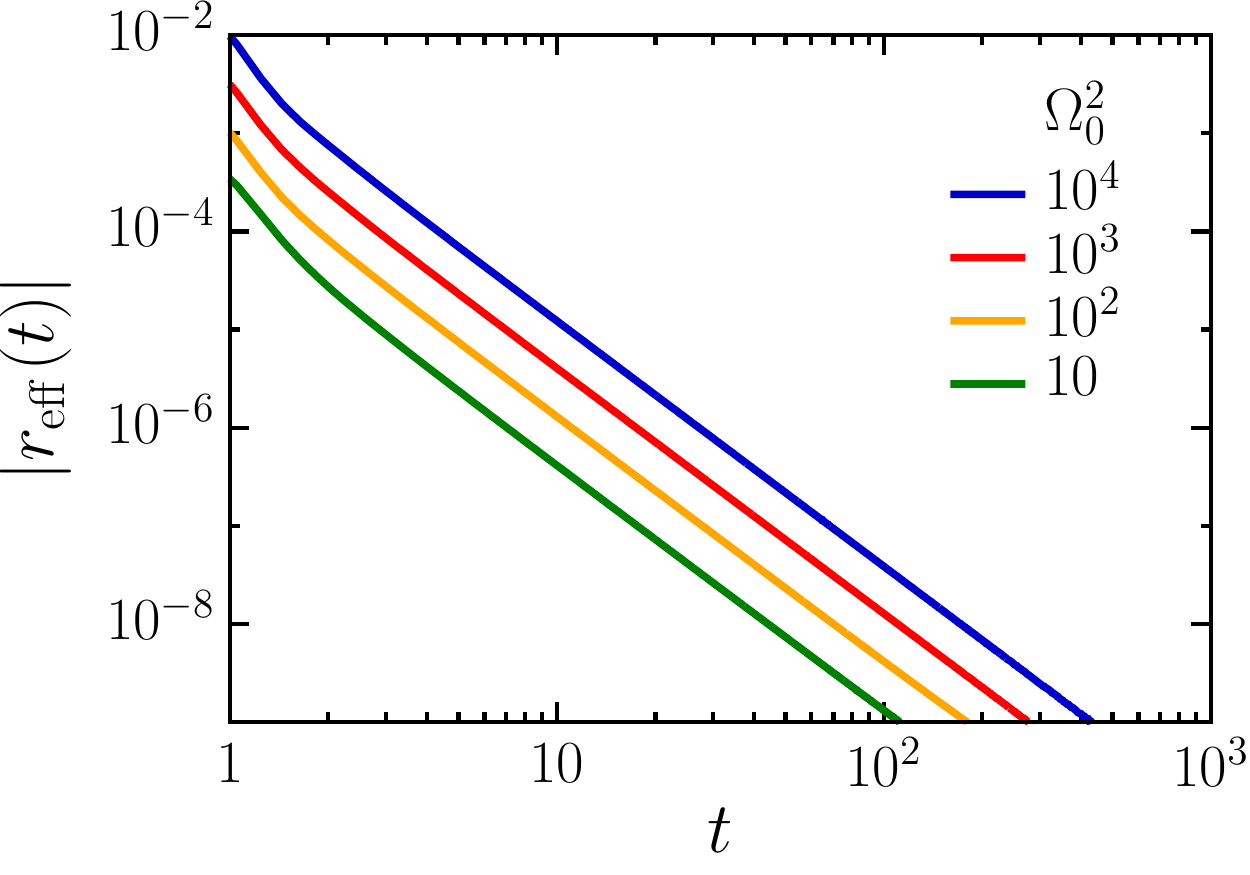}
\end{tabular}
}
\caption{(Color online). Effective parameter $r_{\text{eff}}$ as a function of time $t$.
Upper left panel: $r_{\text{eff}}$ for a quench above (uppermost curve), at (lowermost curve), and below (intermediate curve) the critical point $r_c$ with $d=3.4$, $\Omega_0^2=10^2$, $u \simeq 80$, and a Gaussian cut-off function $h(k/\Lambda) = \exp[-k^2/(2\Lambda^2)]$ with $\Lambda=\pi/2$. The black dashed lines indicate an algebraic decay $\sim t^{-2}$.
Upper central panel: $r_{\text{eff}}$  at $r=r_c$ for {sharp (uppermost curve) and smooth (lowermost curves)} cut-off functions $h(k/\Lambda)$ indicated in the legend (see
also the main text) with $d=3$, $\Omega_0^2=10^2$, $u \simeq 47$, and $\Lambda=\pi/2$.
Upper right panel: $r_{\text{eff}}$  at $r=r_c$  for various values of the sharp cut-off $\Lambda$,{increasing from top to bottom}, with $d=3$, $\Omega_0^2=10^4$, and $u \simeq 4.74$, compared with the expected algebraic decay $\sim t^{-2}$ (dashed line).
Lower left panel: $r_{\text{eff}}$ at $r=r_c$ for various values of the initial parameter $\Omega_0^2$, {decreasing from top to bottom,} with $d=3$, $u \simeq 4.74$, and a Gaussian cut-off function $h$ with $\Lambda = \pi/2$.
Lower central panel: $r_{\text{eff}}$ at $r=r_c$ for several values of $d>d_c=4$, $u=1$, $\Omega_0^2=10^4$ and a Gaussian cut-off function $h$ with $\Lambda = \pi/2$; the dashed lines superimposed on the curves are proportional to $t^{-2.1}$ (uppermost curve), $t^{-2.3}$ (intermediate curve) and $t^{-2.5}$ (lowermost curve).
Lower right panel: $r_{\text{eff}}$ at $r=r_c$ for several values of $\Omega_0$,{decreasing from top to bottom}, with fixed $d=4.5$, $u = 1$ and a Gaussian cut-off function $h$ with $\Lambda = \pi/2$.
}
\label{fig:fig1}
\end{figure*}
%
In Fig.~\ref{fig:fig1} we show the time dependence of $r_\text{eff}(t)$  after a deep quench. In particular, the upper left panel demonstrates, in spatial dimension $d=3.4$, that while $r_\text{eff}(t)$ at long times approaches a finite value for $r>r_c$, it generically vanishes for both $r=r_c$ and  $r < r_c$. In particular, the corresponding decay turns out to be  $\sim t^{-2}$ in both cases, as indicated by the dashed lines, in agreement with the ansatz \eqref{eq:ansatz1} at criticality. Such an algebraic behavior actually sets in after some time $t_\Lambda \sim \mathcal{O}(\Lambda^{-1})$, where $\Lambda$ is the cut-off employed in the algorithm. As expected, the actual possibility of detecting this algebraic decay depends on the way the model is regularized, i.e., on the specific function $h(x)$ used in Eq.~\eqref{eq:self-con1} in order to introduce the cut-off $\Lambda$.
This is illustrated by the upper central panel of Fig.~\ref{fig:fig1} for a quench at criticality $r=r_c$ in $d=3$, in which the momentum integral in Eq.~\eqref{eq:self-con1}
is regularized with a sharp cut-off
as in Eq.~\eqref{eq:def-h-sharp}
(uppermost curve) or with a smooth exponential $h(x) = {\rm e}^{-x}$ or Gaussian $h(x)={\rm e}^{-x^2/2}$ functions (lower curves), with characteristic scale $\Lambda$.
(Note that the value of $r_c$ is also affected by this choice, according to Eq.~\eqref{eq:rc-ansatz}.) While the sharp cut-off causes persistent oscillations in $|r_\text{eff}(t)|$, which mask the expected behavior $\sim t^{-2}$, the smooth ones are qualitatively similar and they reveal this algebraic decay after some time $t_\Lambda$.
The persistent oscillations displayed for a sharp cut-off function are expected to be sub-leading compared to the $\Lambda$-independent decay $\sim t^{-2}$ in a formal expansion in decreasing powers of $\Lambda$: accordingly, their amplitude is expected to decrease as $\Lambda$ increases. This is clearly demonstrated by the curves in the upper right panel of Fig.~\ref{fig:fig1}, which, from top to bottom, slowly approach the expected algebraic behavior (dashed line) upon increasing the value of $\Lambda$. Accordingly, in order to detect the universal behavior $\sim t^{-2}$ in the presence of a sharp cut-off, very large values of $\Lambda$ have to be used, resulting in a longer computational time with respect to that required by an exponential or Gaussian cut-off function $h$.
The onset of a scaling regime for $r_\text{eff}(t)$ for a quench at criticality is also expected to be influenced by the value of the pre-quench parameter $\Omega_0^2$, as discussed in Sec.~\ref{sec:scaling} and in Ref.~\cite{Chiocchetta2015}.  In particular, while the analytic investigation in Sec.~\ref{sec:scaling} assumes a deep quench, i.e., $\Omega_0\gg \Lambda$, it is instructive to check numerically how the actual value of $\Omega_0$ influences the time $t_\Lambda$ after which the expected universal algebraic behavior $\sim t^{-2}$ sets in.
The curves in the lower left panel of Fig.~\ref{fig:fig1} show, from top to bottom, that $t_\Lambda$ increases significantly upon decreasing the value of $\Omega_0^2$, until  the eventual algebraic behavior is completely masked by the initial non-universal transient occurring at $t< t_\Lambda$ when $\Omega_0 \ll \Lambda$.
If the spatial dimensionality $d$ of the model is larger than the upper critical dimensionality $d_c=4$, the leading-order temporal decay of  $r_\text{eff}(t)$ is no longer proportional to $t^{-2}$, because the corresponding proportionality constant $a$ vanishes (see Eq.~\eqref{eq:ansatz1}).
In this case, $r_\text{eff}(t)$ still vanishes at long times with the algebraic law $\sim t^{-(d-2)}$ given in Eq.~\eqref{eq:reff-above-ucd}. This dependence is shown in the lower central panel for $d = 4.1$, 4.3 and 4.5.  The theoretical prediction for the exponent of this decay is indicated by the corresponding dashed lines, while its prefactor is a non-universal constant which depends --- in contrast to the case $d<4$ --- on the actual values of the parameters of the system, e.g., $\Omega_0$, as shown in the lower right panel for $d=4.5$.

\begin{figure*}[h]
\centerline{
	\begin{tabular}{cc}
	\includegraphics[width=6cm]{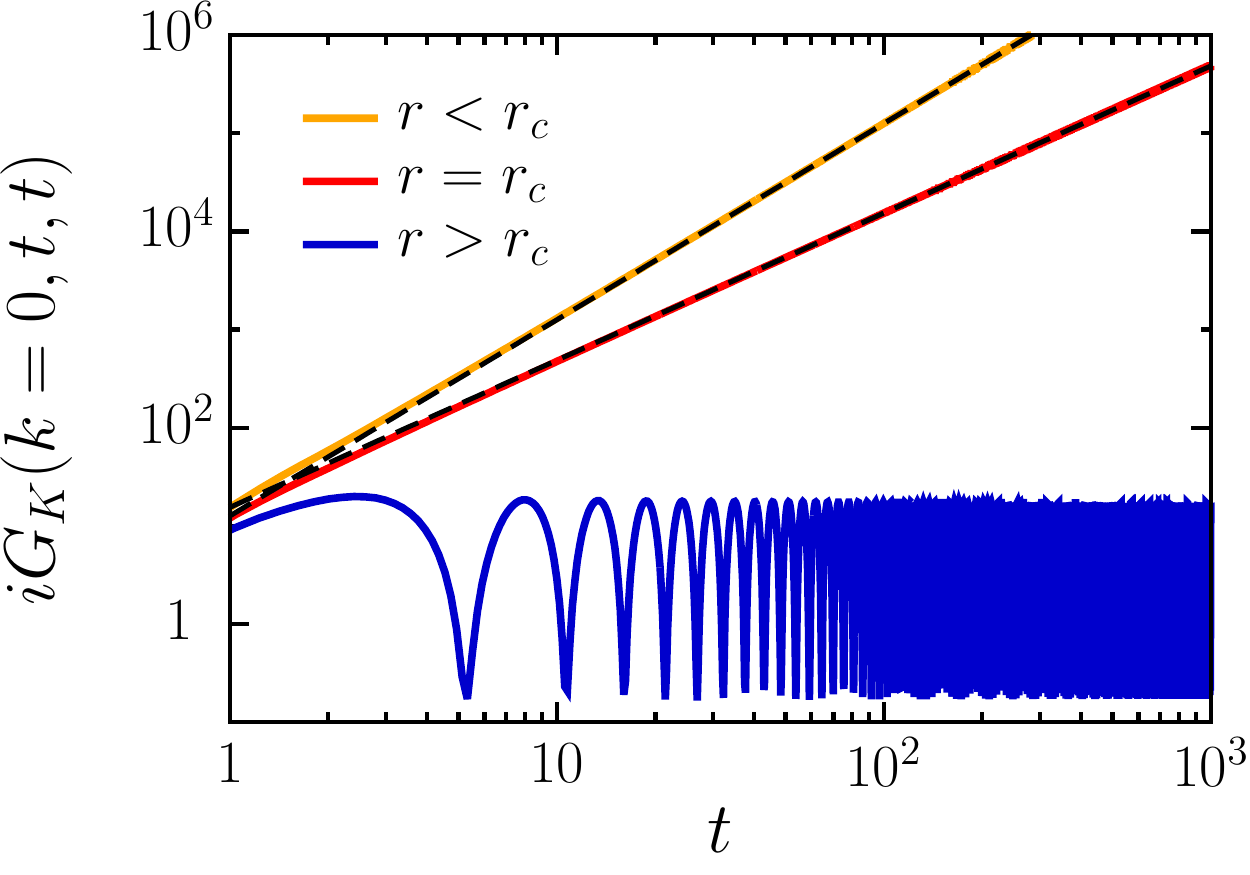} &
	\includegraphics[width=6cm]{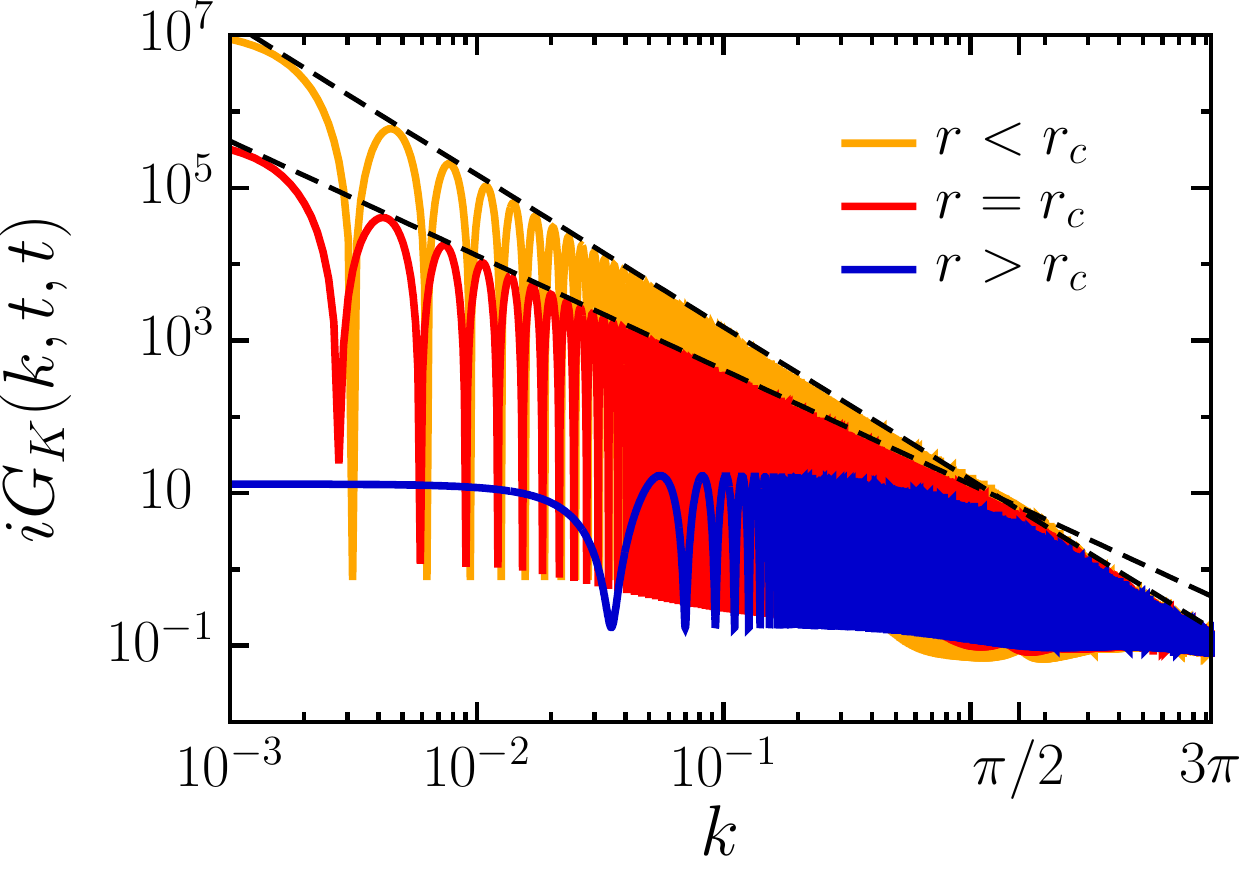} \\
	\includegraphics[width=6cm]{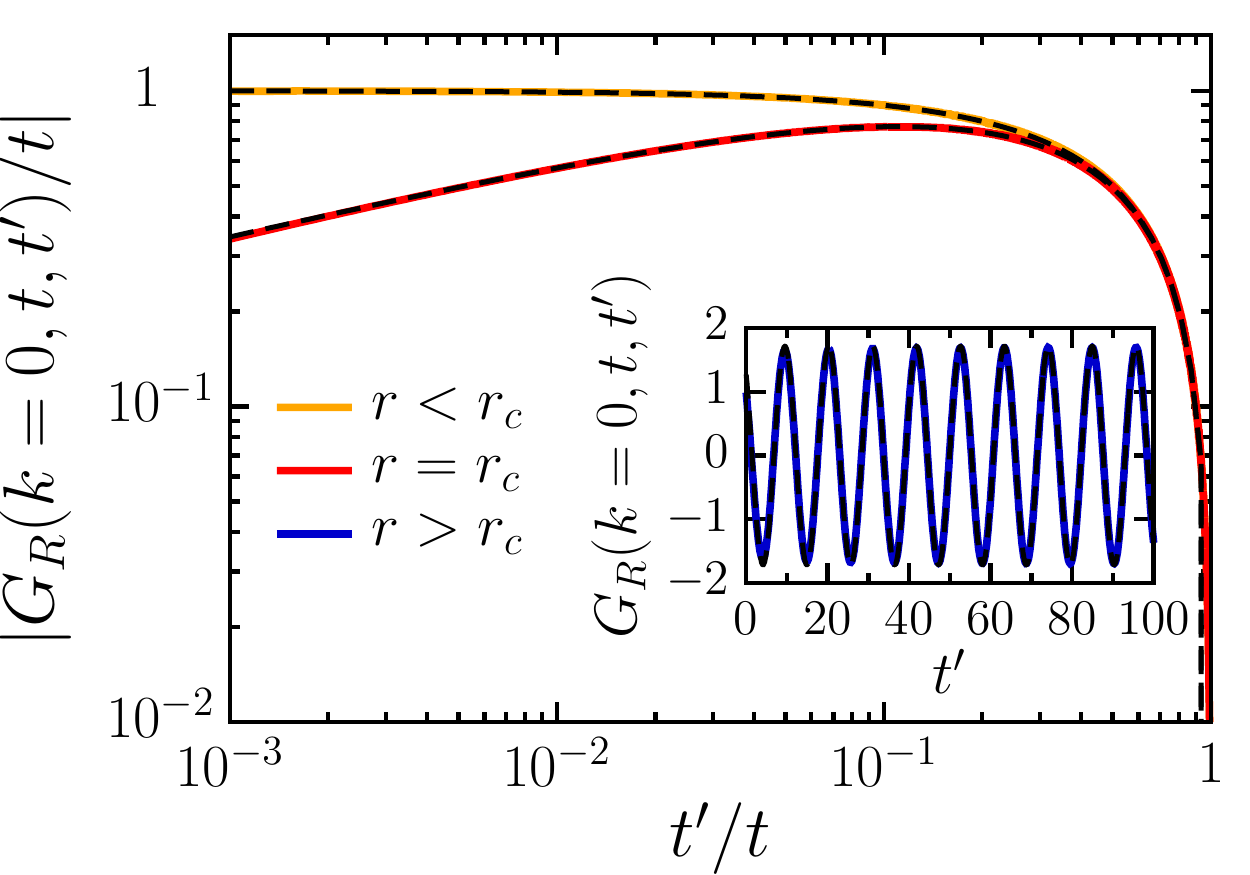} &
	\includegraphics[width=6cm]{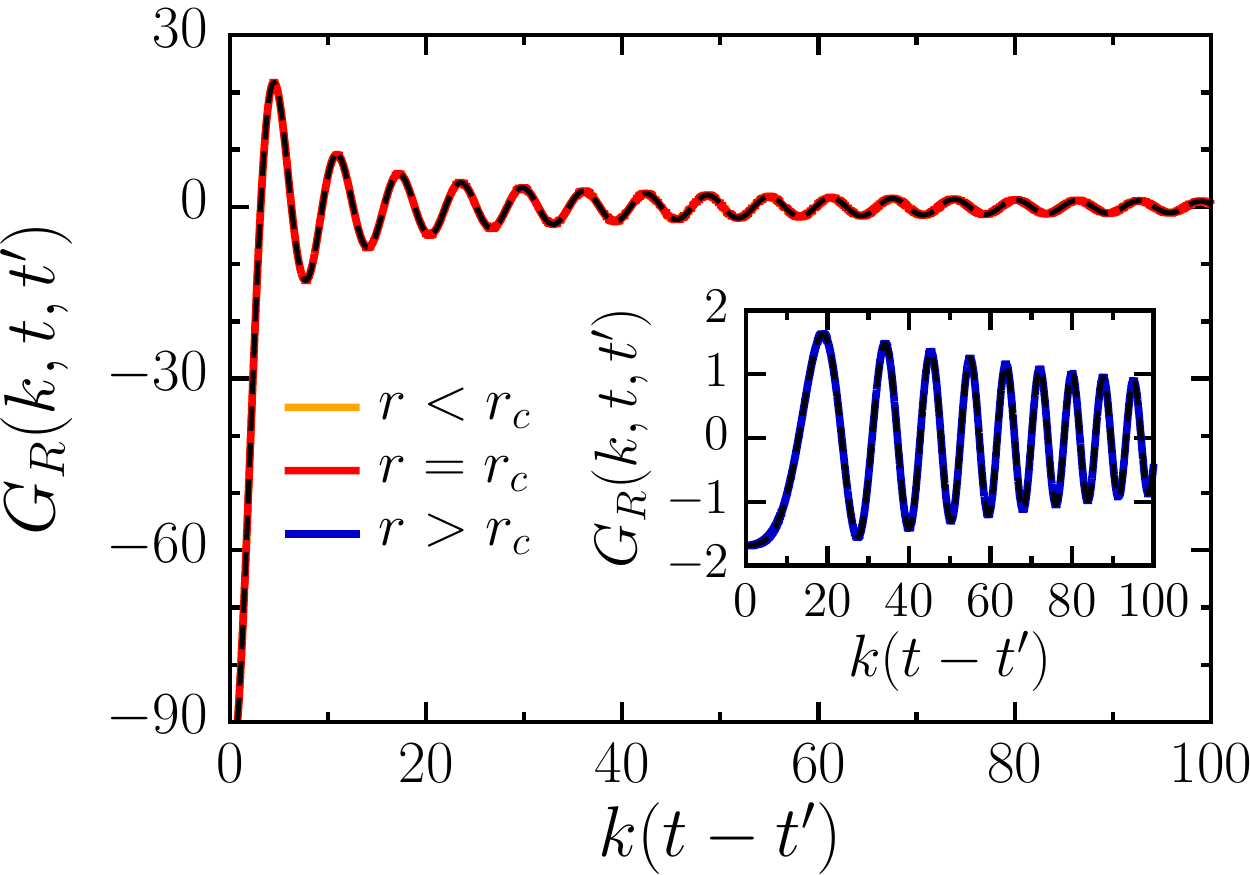}
\end{tabular}
}
\caption{(Color online). Keldysh and retarded Green's functions $G_K$ and $G_R$, respectively.
Upper left panel: $G_K(k,t,t)$ at $k=0$ and equal times as a function of time $t$, for $r$ below (uppermost curve), at (intermediate curve), and above (lowermost curve) the critical value $r_c$. The dashed lines superimposed to the curves for $r=r_c$ and $r<r_c$ are proportional  to $t^{3/2}$ and $t^2$, respectively.
Upper right panel: $G_K(k,t,t)$ at fixed equal times $t=10^3$ as a function of momentum $k$ after a quench with $r$ below {(yellow, uppermost line), at (red, intermediate line) and above (blue, lowermost line)} the critical value $r_c$. The dashed lines superimposed on the curves for $r=r_c$ and $r<r_c$ are proportional to $k^{-3/2}$ and $k^{-2}$, respectively.
Lower left panel, main plot: retarded Green's function $\left| G_R(k, t, t')/t \right|$ for $k=0$ and $t=2\times 10^3$ as a function of $t'/t$ for quenches of $r$ at (lower curve) and below (upper curve) the critical value $r_c$. The black dashed lines superimposed on the curves for $r=r_c$ and $r<r_c$ correspond to  the theoretical predictions
$2 \Phi_{1/4}(t'/t)$ (see Eq.~\eqref{eq:fit-sf}) 
and $\Phi_{1/2}(t'/t)$, respectively.
Inset: retarded Green's function $G_R(k, t, t')$ for $k=0$ and $t=2\times 10^3$ as a function of $t'$ for quenches of $r$ above the critical value $r_c$. The black dashed line corresponds to the prediction in Eq.~\eqref{eq:GR-long-time} with $k \to \sqrt{r^*}$ (see the main text) and $r^*\simeq 0.34$.
Lower right panel: retarded Green's function $G_R(k,t,t')$ as a function of $k(t-t')$ for $t=2\times 10^3$, $t' = 1.9 \times 10^3$ and $r$ below {(main plot, yellow, superimposed on the red one), at (main plot, red, superimposed on the yellow one) and above (inset, blue, solid curve)} the critical value $r_c$. In the main plot, the two curves are indistinguishable from the superimposed dashed line, which corresponds  to the theoretical prediction in Eq.~\eqref{eq:GR-long-time}. In the inset, the curve is indistinguishable from the superimposed dashed line which corresponds 
to Eqs.~\eqref{eq:GR-long-time} and \eqref{eq:sub-k} with $r^* \simeq 0.34$.
In all the panels of this figure, the spatial dimensionality is fixed to  $d=3$, $\Omega_0^2=10^2$, $u \simeq 47$, and a Gaussian cut-off is used with $\Lambda = \pi/2$.
}
\label{fig:fig2}
\end{figure*}
%
As discussed in Sec.~\ref{sec:scaling}, the scaling behavior of $r_\text{eff}(t)$ implies the emergence of algebraic dependences on time in $G_{R,K}$, numerical evidences for which are presented in Fig.~\ref{fig:fig2} for a deep quench occurring at $d=3$.
In particular, $G_K(k,t,t)$ at criticality is expected to display the short-time scaling in Eq.~\eqref{eq:GK-short-time}  for $t \ll k^{-1}$ (which actually extends to long times for $k=0$), while in the long-time limit $t \gg k^{-1}$, $G_K$
displays (up to oscillating terms) the scaling in Eq.~\eqref{eq:GK-long-time} as a function of $k$. In order to corroborate these predictions, the left and right upper panels of Fig.~\ref{fig:fig2} show the dependence on time $t$ and momentum $k$, respectively, of $G_K$ within these two regimes, for quenches occurring at $r>r_c$ (lowermost curves), $r=r_c$ (intermediate curves), and $r<r_c$ (uppermost curves). The upper left panel demonstrates that $G_K(k=0,t,t)$, after a quench to criticality,  grows in time as $\sim t^{3/2}$ (intermediate dashed line) in agreement with Eqs.~\eqref{eq:GK-short-time} and \eqref{eq:a-analytic} for $d=3$. If the quench occurs above criticality, instead, $G_K$ displays oscillations with an asymptotic period $\propto \xi = (r^*)^{-1/2}$ where $r^* = r_\text{eff}(t\to \infty)$
as expected on general grounds~\cite{Chiocchetta2015} and as suggested by the lowermost curve.
Remarkably, an algebraic behavior $\sim t^2$ (uppermost dashed line)  emerges  for quenches occurring below the critical point (i.e., with $r<r_c$), which signals that the corresponding coarsening occurs on a dynamical length scale which grows in time~\cite{Bray1994}.
Further below (see Fig.~\ref{fig:fig4}),  we discuss the dependence of the power of this algebraic decay on the spatial dimensionality $d$. We anticipate here that the present law $\sim t^2$ agrees with what one would obtain by extending the scaling prediction in Eq.~\eqref{eq:GK-short-time} to quenches below $r_c$ and by using the value $\alpha_\text{co} = 1/2$ from Eq.~\eqref{eq:exp-co} instead of $\alpha$.
The upper right panel, instead, reports the dependence on $k$ of the value eventually reached by $G_K(k,t,t)$ at a fixed but long time $t \gg k^{-1}$.
In particular, at criticality $r=r_c$ (intermediate curve), $G_K$ approaches, up to oscillatory terms, the algebraic behavior $\sim k^{-3/2}$ (lower dashed line) in agreement with Eqs.~\eqref{eq:GK-long-time} and \eqref{eq:a-analytic} in $d=3$.
For $r<r_c$ (uppermost curve), $G_K$ still displays, up to oscillatory terms, an algebraic dependence on $k$, but with a different power $\sim k^{-2}$ (upper dashed line) which again agrees with the extension of the critical scaling form \eqref{eq:GK-long-time} below $r_c$ with $\alpha$ replaced by $\alpha_\text{co}$.
When the quench occurs, instead, above the critical point, $G_K$ tends to a constant, up to oscillations. In all the cases illustrated in Fig.~\ref{fig:fig2}, non-universal contributions affect the various curves for $k \gtrsim \Lambda = \pi/2$, due to the effects of the regularizing function $h$.
As far as $G_R(k,t,t')$ is concerned, Eq.~\eqref{eq:GR-fin} provides its complete expression within the scaling regime at criticality $r=r_c$. In particular, $G_R(k=0,t,t')$ acquires the scaling form \eqref{eq:GR-short-time}, while $G_R(k,t,t')$  becomes time-translationally invariant at long times $k^{-1}\ll t'<t$ as in Eq.~\eqref{eq:GR-long-time}. The lower left panel of Fig.~\ref{fig:fig2} shows that $G_R(k=0,t,t')/t$ with $t'<t$ becomes indeed a function of the ratio $t'/t$ only. At criticality (lowermost curve) this agrees with what is expected on the basis of Eqs.~\eqref{eq:GR-short-time} and \eqref{eq:a-analytic} in $d=3$  (which renders $\alpha=1/4$), with $|G_R(k=0,t,t')/t| \propto \Phi_{1/4}(t'/t)$ (dashed line), where we define
\begin{equation}
\label{eq:fit-sf}
\Phi_\alpha(x) \equiv  x^{1/2-\alpha} - x^{1/2+\alpha}.
\end{equation}
For a quench below $r_c$, instead, the same quantity becomes $|G_R(k=0,t,t')/t| \propto 1- (t'/t)$ (dashed line); this agrees with $\Phi_{1/2}(t'/t)$, i.e., with what one would infer by extending the critical scaling function \eqref{eq:GR-short-time} below $r_c$ and by using the value $\alpha_\text{co}=1/2$ in Eq.~\eqref{eq:exp-co} for the exponent $\alpha$.
The inset shows, instead, the numerical data for $G_R(k=0,t,t')$ after a quench to $r>r_c$, as a function of $t'$ and fixed $t>t'$, which is characterized by persistent oscillations.
As mentioned above, for this kind of quench, $r_\text{eff}(t)$ approaches a finite asymptotic value $r^*$ at large times and, up to the leading order, the system behaves as a Gaussian model quenched at $r^*$, for which the response function is given by~\cite{Chiocchetta2015} Eq.~\eqref{eq:GR-long-time} with 
\begin{equation}
k \to \sqrt{k^2 + r^*}.
\label{eq:sub-k}
\end{equation}
The dashed line reported in the inset of the figure, which is actually indistinguishable from the numerical data, corresponds to this theoretical prediction with $k=0$ (see also next panel), confirming its accuracy.
The lower right panel of Fig.~\ref{fig:fig2} shows $G_R(k,t>t',t')$ as a function of $k(t-t')$ for two fixed long times $t$ and $t'<t$ and upon varying $k$.
The main plot shows the corresponding numerical curves both for $r=r_c$ and $r<r_c$, which are however perfectly superimposed and practically  indistinguishable from the theoretical prediction in Eq.~\eqref{eq:GR-long-time} (dashed line), the latter being independent of the actual value of  $\alpha$.
The inset, instead, shows $G_R$ for $r>r_c$: also in this case, the numerical data are indistinguishable from the corresponding theoretical prediction obtained on the basis of 
Eqs.~\eqref{eq:GR-long-time} and \eqref{eq:sub-k},
as explained above while illustrating the previous panel.

\begin{figure*}[h]
\centerline{
	\begin{tabular}{cc}
	\includegraphics[width=6cm]{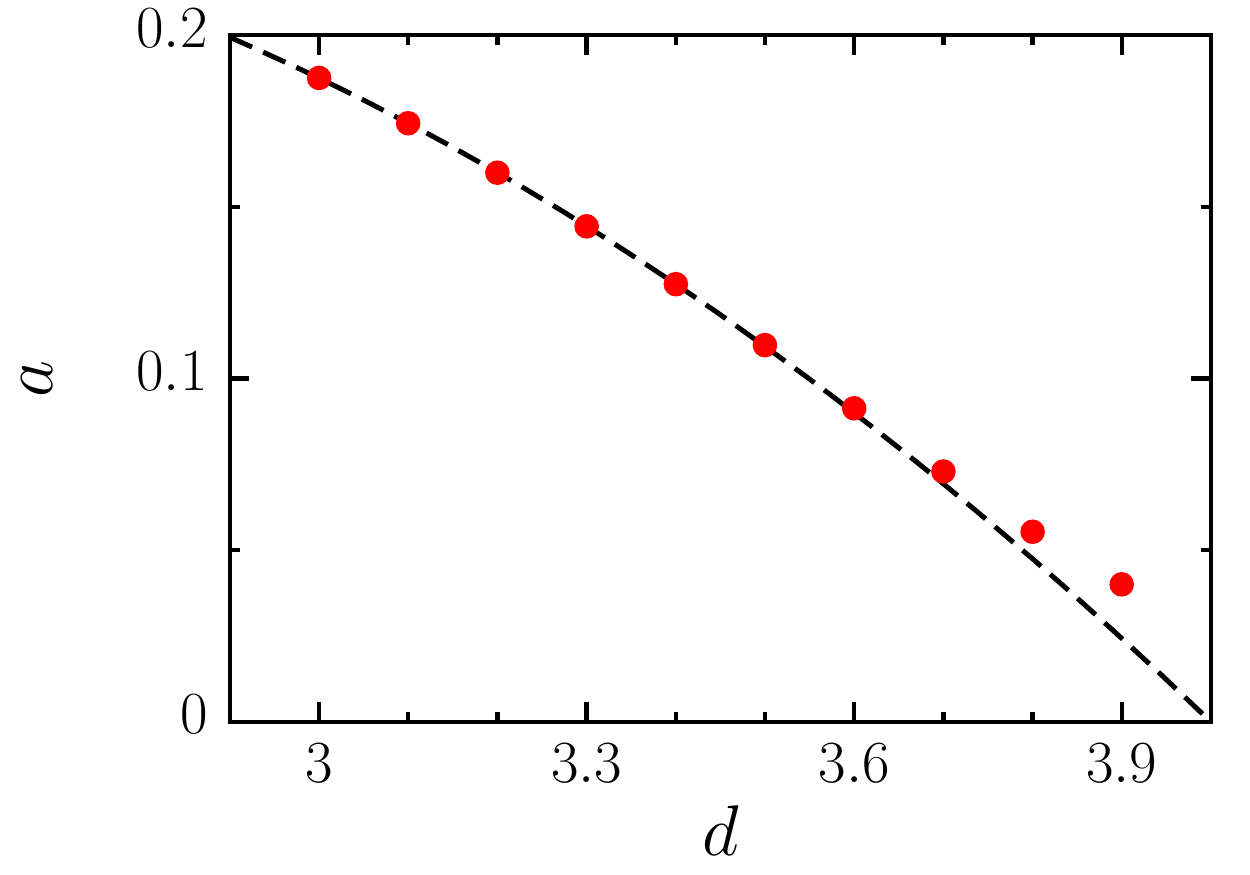} &
	\includegraphics[width=6cm]{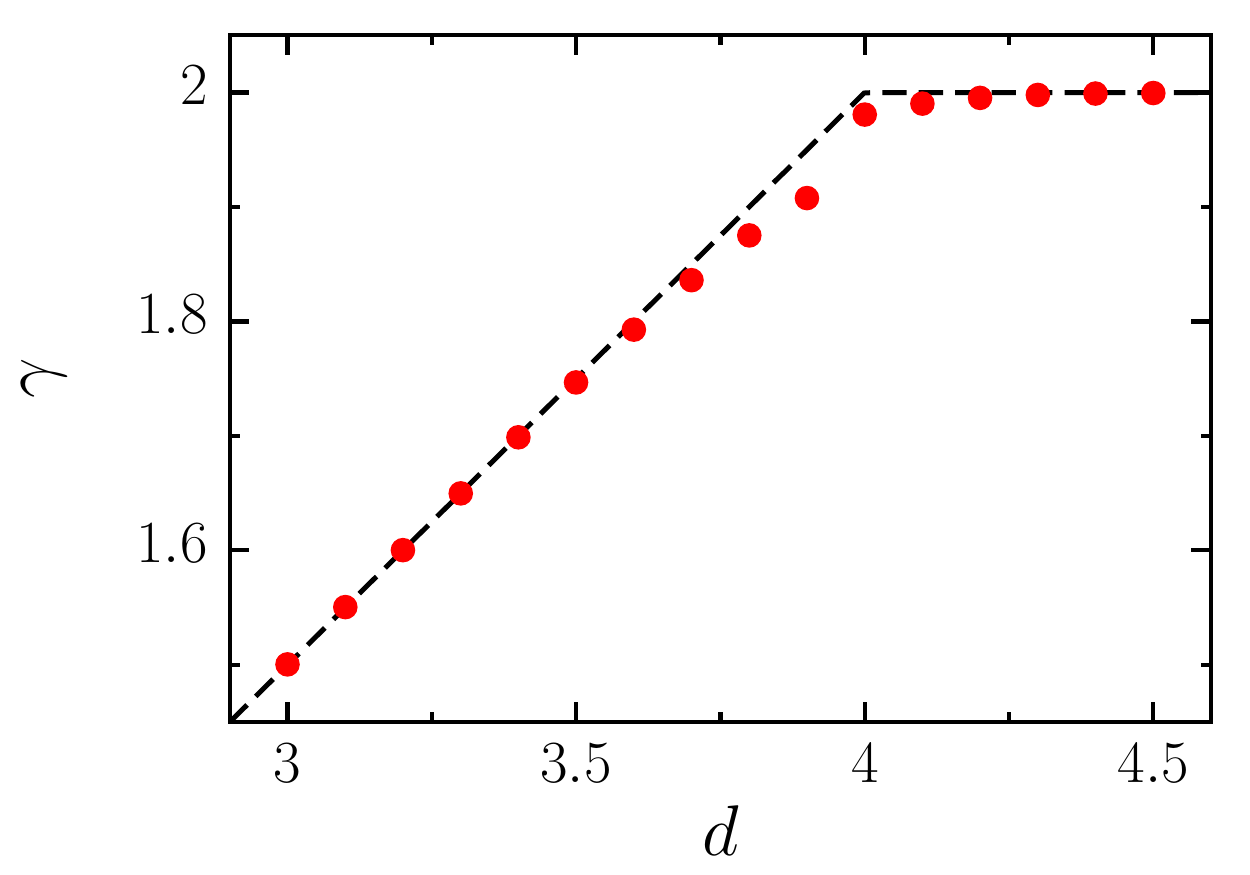} \\
	\includegraphics[width=6cm]{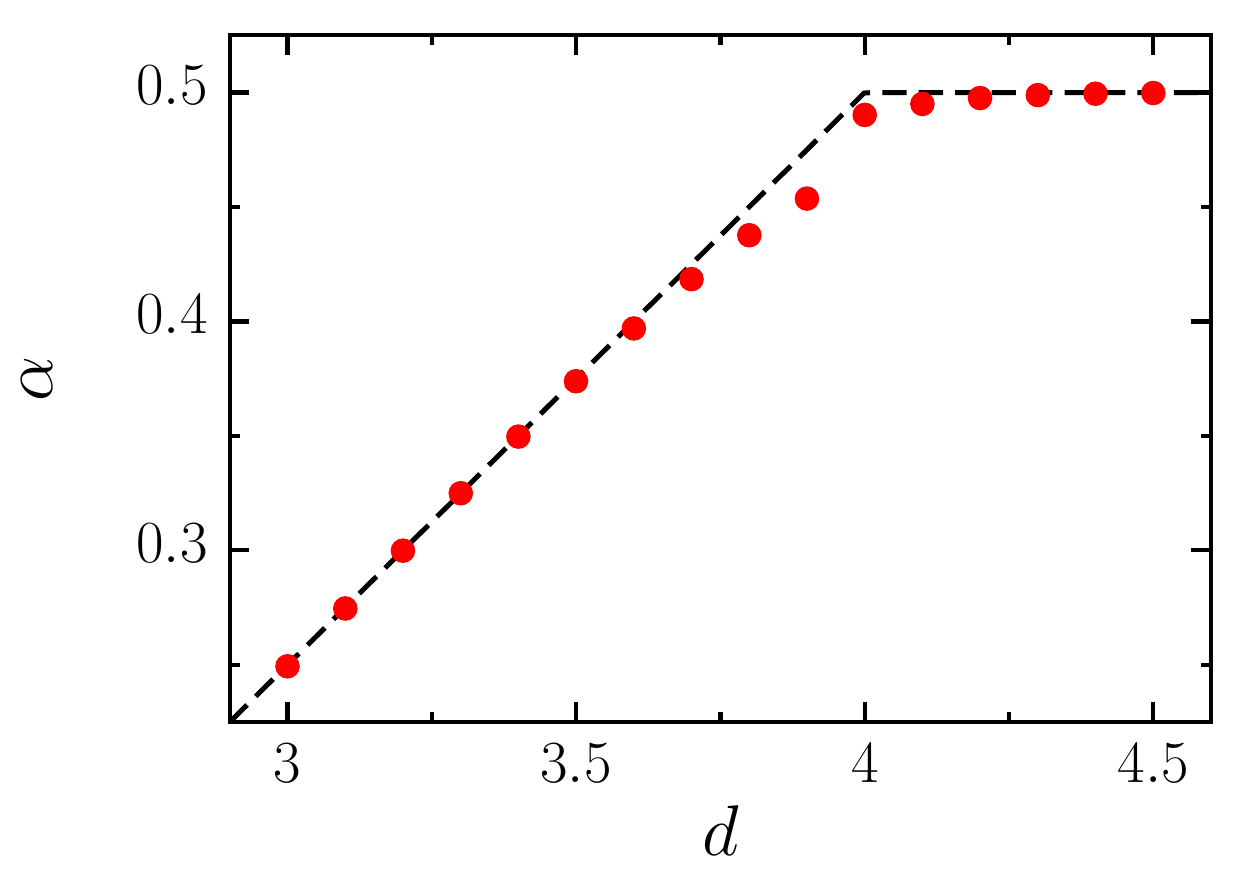} &
	\includegraphics[width=6cm]{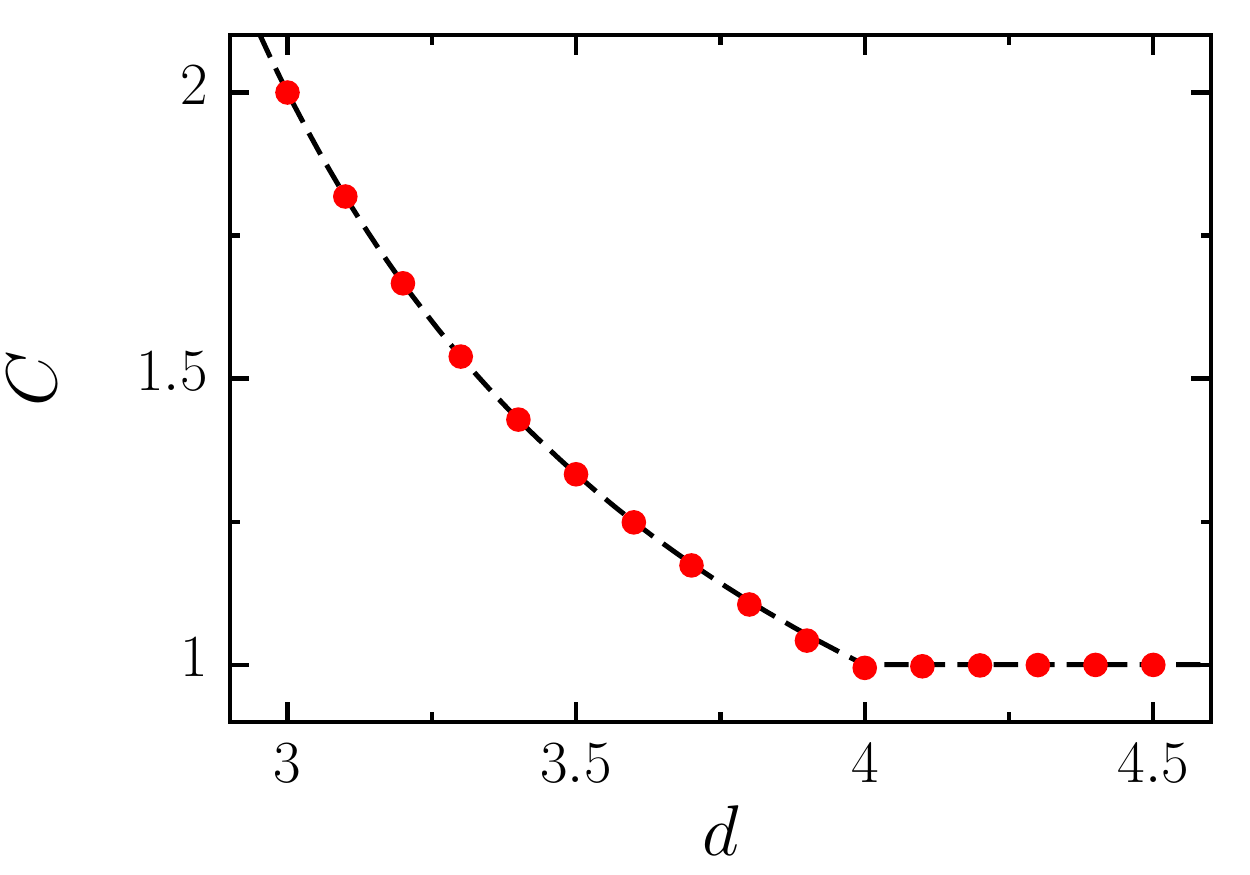}
\end{tabular}
}
\caption{(Color online). Numerical values (symbols) of the exponents $\gamma$, and $\alpha$ and prefactors $a$ and $C$ as functions of the spatial dimensionality $d$ for a quench at the critical point.
Upper left panel: coefficient $a$ of the effective parameter $r_\text{eff}(t)$ computed numerically compared with the theoretical value $a = d(1-d/4)/4$ (dashed line).
Upper right panel: exponent $\gamma$ obtained by fitting $i G_K(k=0, t, t)$ with an algebraic law $\propto t^{\gamma}$; the dashed line indicates the analytical value $\gamma = d/2$ for $d<4$ and $\gamma=2$ for $d>4$.
Lower left panel: exponent $\alpha$ obtained by fitting $|G_R(k=0, t=2\times 10^3, t')/t|$ with $C\, \Phi_\alpha(t'/t)$, see Eq.~\eqref{eq:fit-sf}; the dashed line represents the theoretical value $\alpha = (d-2)/4$ for $d<4$ and $\alpha = 1/2$ for $d>4$.
Lower right panel: prefactor $C$ of $G_R(k=0, t, t')$ as a function of the dimension $d$, obtained by fitting $|G_R(k=0, t=2\times 10^3, t')/t|$ with $C  \Phi_\alpha(t'/t)$, with $\alpha$ given by the theoretical values reported above; the dashed line represents the theoretical prediction for $C = 2/(d-2)$ for $d<4$ and $C=1$ for $d\ge 4$ (see Eq.~\eqref{eq:GR-short-time}).
The numerical data presented in this figure have been obtained with a Gaussian cut-off function with $\Lambda = \pi/2$ and $\Omega_0^2=10^4$, while for each point a different (inconsequential) value of $u$ has been used. The statistical error bars on the numerical points are smaller than the symbol size.}
\label{fig:fig3}
\end{figure*}
%
The numerical results presented in Figs.~\ref{fig:fig1} and \ref{fig:fig2} refer to a quench of the model in a certain spatial dimensionality $d$. In order to test both the predictions in Eq.~\eqref{eq:a-analytic} and some of the features of the scaling functions at criticality we repeated the analysis of the previous figures for a variety of values of $d$, the results of which are reported in Fig.~\ref{fig:fig3}.
In particular,
the long-time behavior of $r_\text{eff}(t)$ after a quench at $r=r_c$ (such as the one displayed for $d=3.4$ in the upper left panel of Fig.~\ref{fig:fig1}) can be fitted with  the expected algebraic law $a t^{-2}$ in order to extract the value of $a$ as a function of $d$. The resulting numerical estimates are indicated by the dots in the upper left panel of the figure, where they are compared with the theoretical prediction in Eq.~\eqref{eq:a-analytic} (dashed line). While the agreement between the latter and the numerical data is very good for $d\lesssim 3.6$, slight deviations appear upon approaching the upper critical dimensionality $d_c=4$ of the model, due to the expected corrections to scaling which are known to become increasingly relevant as $d\to d_c$~\cite{ZinnJustinbook}.

Analogously, by fitting the time dependence of the critical $iG_K(k=0,t,t)$ with the algebraic law $\sim t^\gamma$, one can estimate the numerical value of $\gamma$ as a function of $d$. These estimates, reported in the upper right panel of Fig.~\ref{fig:fig3} (symbols), can then be compared with the analytical prediction $\gamma \equiv 1 + 2 \alpha = d/2$ for $d<4$ and $\gamma =2$ for $d\ge 4$ (dashed line) which follows from Eqs.~\eqref{eq:GK-short-time}, \eqref{eq:a-analytic}, and \eqref{eq:a-analytic-mf}. Also in this case the agreement between the data and the analytical prediction is very good apart from a region around $d_c$, where corrections to scaling make the extraction of the exponent from the data more difficult.
While both of the previous evidences in favor of the theoretical predictions of Sec.~\ref{sec:scaling} are based on the scaling properties of $G_K$ and of $r_\text{eff}(t)$, an independent and more stringent test is provided by the analysis of the response function $G_R$, whose analytical form in Eqs.~\eqref{eq:GR-fin} and \eqref{eq:GR-short-time} does not involve unknown parameters (such as those which fix, instead, the amplitude of $G_K(k=0,t,t)$ in Eq.~\eqref{eq:GK-short-time}).
In particular, another estimate of $\alpha$ can be obtained by fitting $|G_R(k=0,t,t')/t|$ with $C \, \Phi_\alpha(t'/t)$ (see Eq.~\eqref{eq:fit-sf}) as predicted by Eq.~\eqref{eq:GR-short-time}. The resulting values of $\alpha$ are reported in the lower left panel of Fig.~\ref{fig:fig3} together with the analytical prediction (dashed line) of Eqs.~\eqref{eq:a-analytic} and \eqref{eq:a-analytic-mf}.  Alternatively, one can estimate the value of the proportionality constant $C$ by fitting $|G_R(k=0,t,t')/t|$ with $C \Phi_\alpha(t'/t)$ where $\alpha$ is now fixed to the theoretically expected value reported in Eqs.~\eqref{eq:a-analytic} and \eqref{eq:a-analytic-mf}. The resulting numerical estimates are indicated by the symbols in the lower right panel of Fig.~\ref{fig:fig3} together with the analytical prediction (dashed line) $C = 1/(2\alpha)$ which follows from Eq.~\eqref{eq:GR-short-time} and from the theoretical values of $\alpha$ of Eqs. \eqref{eq:a-analytic} and \eqref{eq:a-analytic-mf}. Also for the lower left panel, the agreement with theoretical predictions is 
good, except for values close to $d_c$, while it is remarkably good for the lower right panel.
Further below we argue that the exponent $\gamma = 1 + 2\alpha$ which describes the algebraic behavior of $G_K(k,t,t)$ both at short and long times (see Eqs.~\eqref{eq:GK-short-time} and \eqref{eq:GK-long-time}) is the same as the one
introduced in Ref.~\cite{Smacchia2014} in order to characterize the small-momentum behavior $\sim k^{-\gamma}$ of $\rho_k(t)$ up to a cut-off $k^*\sim t^{-1}$. The quantity $\rho_k(t)$ corresponds to the average number of excitations with momentum $k$ of the pre-quench Hamiltonian which are produced after a so-called double quench, i.e., when the parameters of the post-quench Hamiltonian are restored suddenly to their initial values after a time $t$ has elapsed from the first quench occurring at $t=0$. The values of $\gamma$ which were numerically determined in Ref.~\cite{Smacchia2014} for a quench to the critical point in $d=3$ and $4$, i.e., $\gamma = 3/2$ and $2$ are in perfect agreement with our numerical estimates and analytical predictions reported in the upper right panel of Fig.~\ref{fig:fig3}.

\subsection{Quench below the critical point}
\label{sec:coarsening}

%
\begin{figure*}
\centerline{
	\begin{tabular}{cc}
	\includegraphics[width=6cm]{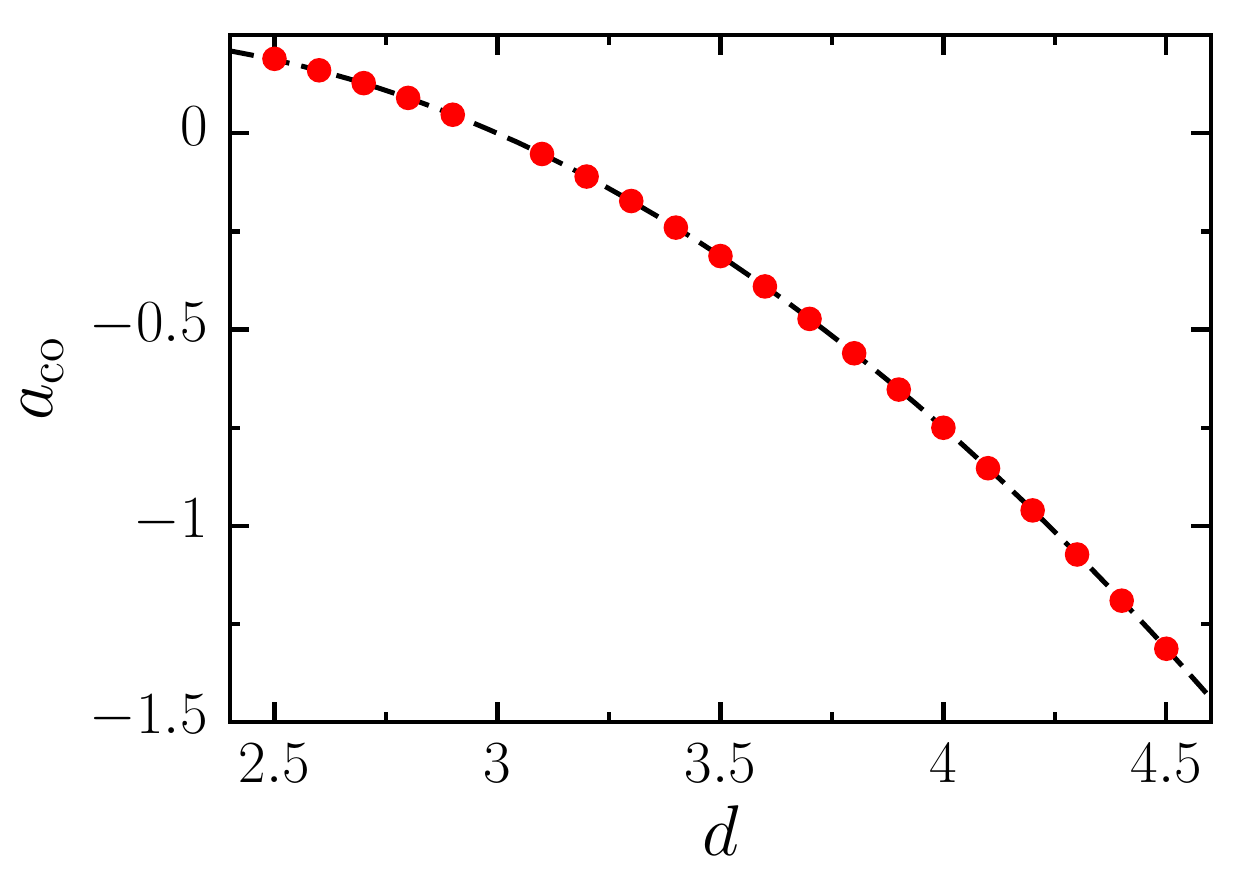} &
	\includegraphics[width=6cm]{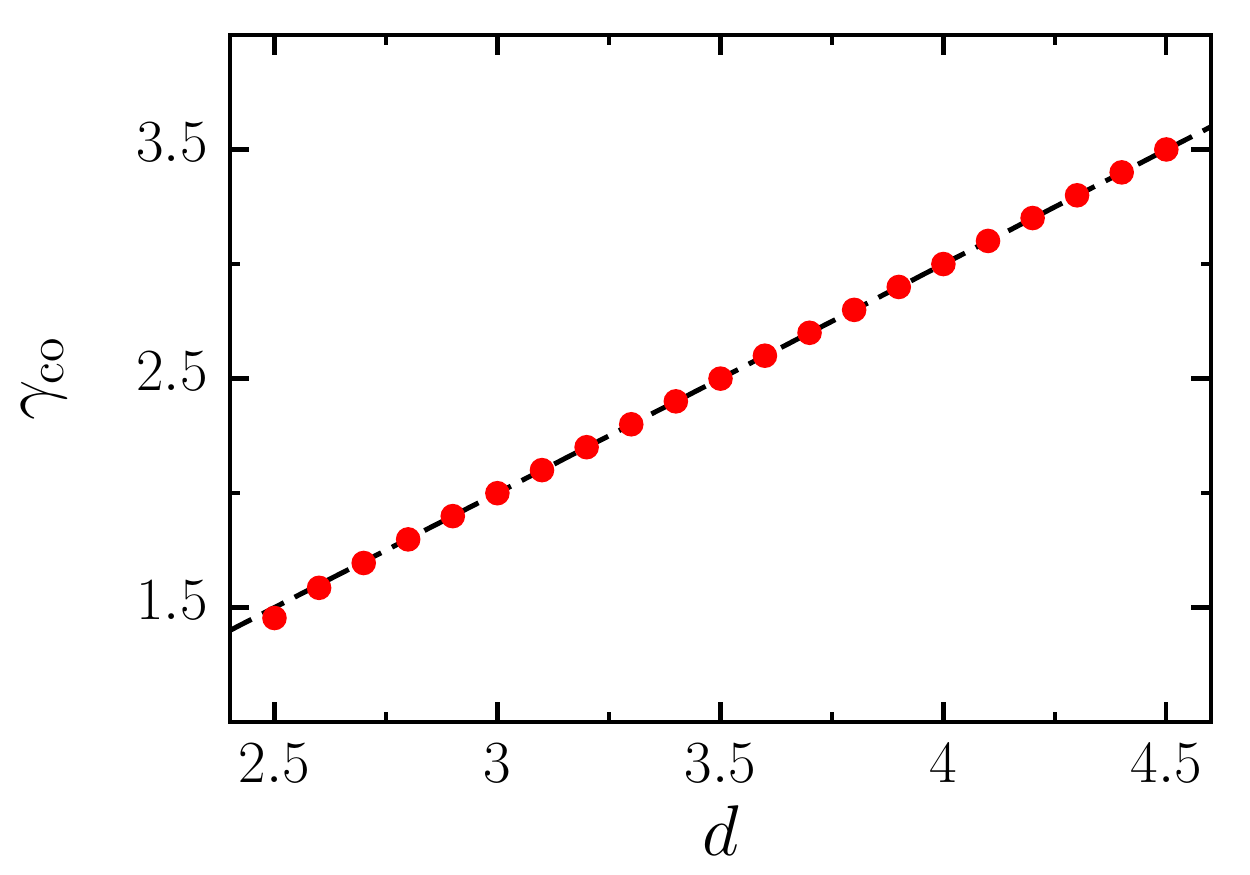} \\
	\includegraphics[width=6cm]{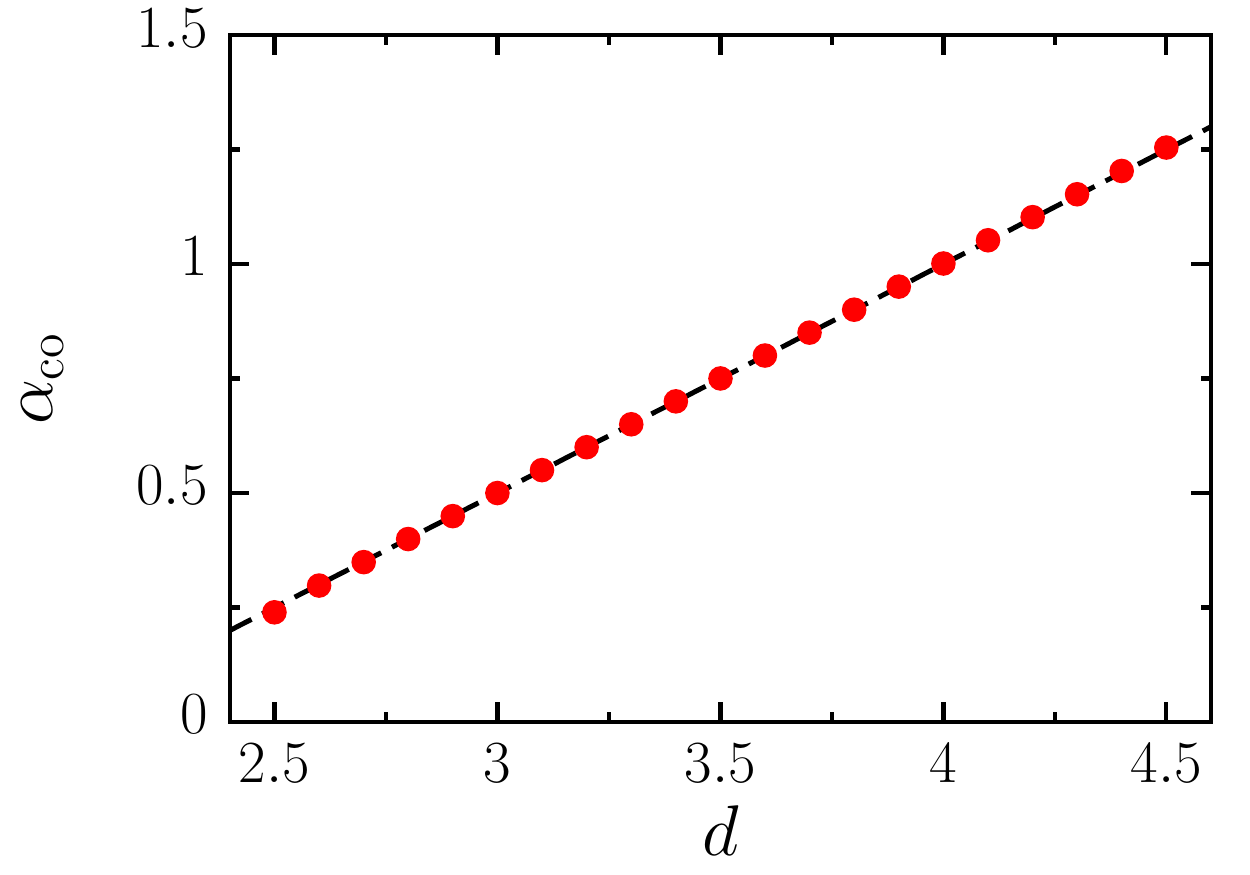} &
	\includegraphics[width=6cm]{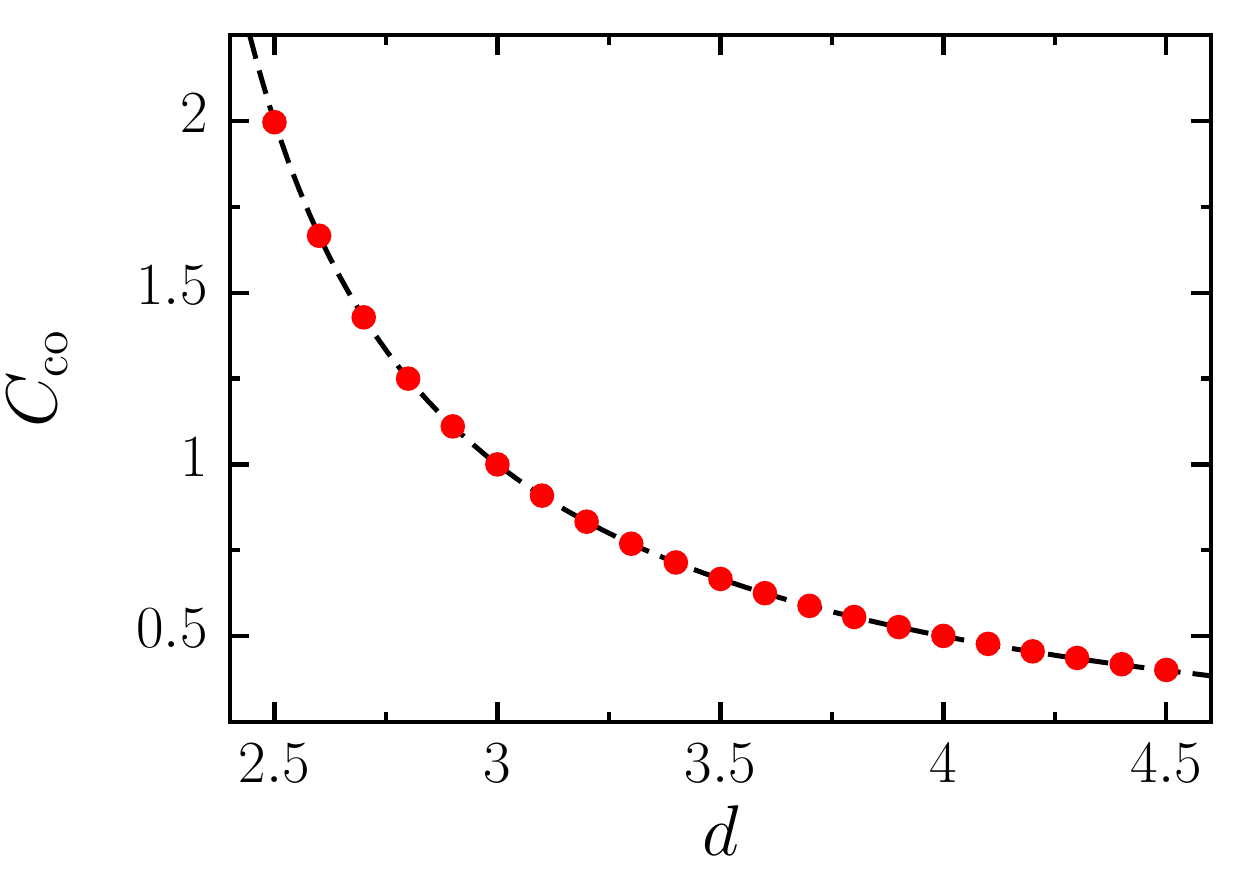}
\end{tabular}
}
\caption{(Color online). Numerical values (symbols) of exponents $\gamma_\text{co}$, and $\alpha_\text{co}$ and prefactors  $a_\text{co}$ and $C_\text{co}$ as functions of dimension $d$ for a quench below the critical point. Upper left panel: numerical estimates  for the coefficient $a_\text{co}$ of the algebraic decay of the effective parameter $r_\text{eff}(t) \simeq a_\text{co} t^{-2}$ as a function of dimension $d$, compared with the theoretical value $a_\text{co} = (3-d)(d-1)/4$ (dashed line) . Upper right panel: numerical estimates of the exponent $\gamma_\text{co}$ obtained by fitting $i G_K(k=0, t, t)$ with $t^{\gamma_\text{co}}$; the dashed line represents the analytical prediction $\gamma_\text{co} = d-1$.
Lower left panel: exponent $\alpha_\text{co}$ obtained by fitting $|G_R(k=0, t=2\times 10^3, t')/t|$ as $C_\text{co}\,\Phi_{\alpha_\text{co}}(t'/t)$ (with $\Phi_\alpha(x)$ given in Eq.~\eqref{eq:fit-sf}); the dashed line represents the theoretical value $\alpha_\text{co} = (d-2)/2$. Lower right panel: prefactor $C_\text{co}$ of $G_R(k, t, t')$ as a function of the dimension $d$, obtained by fitting $|G_R(k=0, t=2\times 10^3, t')/t|$ as $C_\text{co} \Phi_{\alpha_\text{c0}}(t'/t)$, with $\alpha_\text{co}$ fixed to its corresponding theoretical value (see Eq.~\eqref{eq:exp-co});
the dashed line represents the theoretical value $C_\text{co} = 1/(d-2)$.
The numerical data presented in this figure refer to a system with $\Omega_0^2=10^4$, $r = -3$, and a Gaussian cut-off function with $\Lambda = \pi/2$; for each point a different (inconsequential) value of $u$ has been used. The statistical error bars on the numerical points are smaller than the symbol size.
}
\label{fig:fig4}
\end{figure*}
%
The Keldysh (Fig.~\ref{fig:fig2}, upper panels) and the retarded (Fig~\ref{fig:fig2}, lower panels) Green's functions $G_K$ and $G_R$, respectively, for $r>r_c$  are characterized by an oscillatory behavior,  which denotes the presence of a finite length scale $\xi$ in the model, set by the asymptotic value $r^*$ of $r_\text{eff}(t)$.
On the contrary, for $r<r_c$, the effective parameter $r_\text{eff}(t)$ turns out to decay to zero as $\simeq a_\text{co} t^{-2}$ (see the upper left panel of Fig.~\ref{fig:fig1}), i.e., with the same power law as at criticality $r=r_c$: correspondingly,  $G_{R,K}$ exhibit algebraic behaviors, which however differ from the critical ones. 
In fact, it is rather related to the phenomenon of \emph{coarsening} which we discuss further below in Sec.~\ref{sec:coarsening-exp}.
%
%
%
%
%
As we discussed in Sec.~\ref{sec:scaling}, the value of $a_\text{co}$ --- as well as of $a$ for $r=r_c$ --- can in principle be determined in such a way as to satisfy the self-consistent equations \eqref{eq:self-con2} and \eqref{eq:gammas}, which indeed admit two different solutions, reported in Eqs.~\eqref{eq:a-analytic} and \eqref{eq:exp-co}. While the former correctly describes the observed behavior at criticality (see the evidence presented in Fig.~\ref{fig:fig3}),  it is quite natural to expect the latter to describe the other possible scaling behavior, i.e., the one associated with coarsening.
Numerical evidence of this fact is presented in Fig.~\ref{fig:fig4}. In particular, the upper left panel shows the value of $a_\text{co}$ (symbols) for various values of $d$, as inferred by fitting the corresponding numerical data of $r_\text{eff}(t)$ for $r<r_c$ with $a_\text{co}t^{-2}$. The dashed line corresponds to the theoretical prediction reported in Eq.~\eqref{eq:exp-co}.
Although the numerical data reported in Fig.~\ref{fig:fig4} refer to a quench with $r=-3 < r_c$, we have verified that these numerical estimates are not affected by the choice of $r<r_c$.
Note that while $a$ at $r=r_c$ as a function of the dimensionality $d$ shows a marked change in behavior upon crossing the upper critical dimensionality $d_c$ (see the upper left panel of Fig.~\ref{fig:fig3}), being zero above it, this is not the case for $a_\text{co}$.
Analogous consideration holds for the other quantities discussed further below, when compared with the corresponding ones at criticality.
Heuristically this might be expected based on the fact that --- as in the case of classical systems~\cite{Bray1994,Godreche2002}
--- coarsening for $r<r_c$ is generally driven by a different mechanism compared to the one controlling the behavior at $r=r_c$, which is related to critical fluctuations and which is therefore affected upon crossing $d_c$ (see Sec.~\ref{sec:coarsening-exp}).
As a peculiar feature of $a_\text{co}$, we  note that it vanishes for $d=3$.

As argued above and demonstrated by the curves in the upper left panel of Fig.~\ref{fig:fig2}, $G_K(k=0,t,t)$ grows algebraically both at $r=r_c$ and for $r<r_c$, in the latter case as $\sim t^{\gamma_\text{co}}$. The upper right panel of Fig.~\ref{fig:fig4} shows the estimates of $\gamma_\text{co}$ obtained by fitting the numerical data for $G_K(k=0,t,t)$, as a function of the dimensionality $d$. The dashed line corresponds to the theoretical prediction $\gamma_\text{co} = 1 + 2\alpha_\text{co}$ with $\alpha_\text{co}$ given by Eq.~\eqref{eq:exp-co}. 
As it was done in Fig.~\ref{fig:fig3} for $r=r_c$, the lower panels of Fig.~\ref{fig:fig4} consider $G_R(k,t,t')$ at $k=0$. By assuming that the scaling behavior in Eq.~\eqref{eq:GR-short-time} carries over to $r<r_c$, $|G_R(k,t,t')/t|$ is fitted by $C_\text{co} \Phi_{\alpha_\text{co}}(t'/t)$ (with $\Phi_\alpha$ given in Eq.~\eqref{eq:fit-sf}) in order to extract $\alpha_\text{co}$ (lower left panel) or to estimate $C_\text{co}$ once $\alpha_\text{co}$ has been fixed to its theoretical value in Eq.~\eqref{eq:exp-co}. In both panels the corresponding theoretical predictions are reported as dashed lines and, as in the case of the upper panels, the agreement with the numerical data  is excellent, with a hint of slight deviations upon approaching the lower critical dimensionality $d=2$ of this model.
As we mentioned at the end of Sec.~\ref{sec:critical}, the exponent $\gamma_\text{co} = 1 + 2\alpha_\text{co}$ discussed here in connection to the scaling of $G_K(k,t,t)$ (see Eqs.~\eqref{eq:GK-short-time} and \eqref{eq:GK-long-time}) is the same as the exponent $\gamma$ introduced in Ref.~\cite{Smacchia2014} in order to characterize the scaling behavior of $\rho_k(t)$. The values of $\gamma$ which was numerically determined in Ref.~\cite{Smacchia2014} for a quench below the critical point in $d=3$ and $4$, i.e., $\gamma = 2$ and $3$ are in perfect agreement with our numerical estimates and analytical predictions reported in the upper right panel of Fig.~\ref{fig:fig4}.

\subsection{Coarsening}
\label{sec:coarsening-exp}
The numerical data presented in Sec.~\ref{sec:coarsening} clearly show that the non-equilibrium dynamics of the system after a quench to $r<r_c$ 
features
an emerging scaling behavior which we partly rationalized in Sec.~\ref{sec:scaling} and which is characterized by scaling exponents depending on the spatial dimensionality $d$.
As anticipated, these scaling forms are expected to be related to the coarsening dynamics~\cite{Sondhi2013,Sciolla2013}, analogously to what happens in classical systems after a quench below the critical temperature~\cite{Bray1994,Godreche2002,Cugliandolo2015} (see Appendix~\ref{sec:comparison} for the discussion of a specific example). In fact, when a classical system prepared in a disordered state is quenched below the critical temperature, the global symmetry cannot be dynamically broken and, consequently, the order parameter remains zero in average. 
Nevertheless, symmetry is broken locally by the creation of domains within which the order parameter $\varphi$ takes the value characterizing one of the possible different and competing phases.
The average linear extension $L(t)$
of the ordered domains increases with time $t$, until a specific domain possibly prevails over the others, establishing the equilibrium state. 
However, because of such competition, 
$L(t)$ grows algebraically as $L(t) \propto t^{1/z_\text{c}}$, where $z_\text{c}>0$ is an exponent depending on the universal properties of the model, and equilibrium is reached only in an infinite time. Consequently,
%
this lack of an intrinsic length scale in the system
affects the equal-time two-point correlation function 
\begin{equation}
C(\rr,t) = \langle\varphi(\xx+\rr,t)\varphi(\xx,t)\rangle,
\end{equation}
and its spatial Fourier transform $C(k,t)$, 
which, according to the scaling hypothesis~\cite{Bray1994},  
are expected to display the scaling forms
\begin{equation}
\label{eq:dynamical-scaling-hp}
C(\rr, t) = f(r/L(t))\ \mbox{and} \ C(k,t) = [L(t)]^d \widetilde{f}(kL(t)), 
\end{equation}
where $d$ is the spatial dimensionality and $\widetilde{f}(x)$ the Fourier transform of $f(x)$.


%
The scaling forms for a quench below $r_c$ highlighted in Sec.~\ref{sec:coarsening} \emph{do not} satisfy the scaling hypothesis \eqref{eq:dynamical-scaling-hp}, as it was noticed for $d=3$ in 
Refs.~\cite{Sciolla2013,Sondhi2013}. 
In fact, the Keldysh Green's function at equal times $G_K(k,t,t)$ --- which corresponds to the correlation function $C(k,t)$ mentioned above --- can be written as a scaling form by using Eqs.~\eqref{eq:GK-scaling} and \eqref{eq:exp-co}, which reads:
\begin{equation}
\label{eq:coarsening-violation}
G_K(k,t,t) = [L(t)]^{\gamma_\text{co}} \mathcal{G}_d(k L(t)),
\end{equation}
where $\mathcal{G}_d(x)$ is the scaling function, $L(t) \propto t$ (i.e., the coarsening exponent $z_c$  takes the value $z_c=1$) and $\gamma_\text{co}  = d-1$. As this $\gamma_\text{co}$ differs from $d$, Eq.~\eqref{eq:coarsening-violation} violates the scaling form~\eqref{eq:dynamical-scaling-hp} in all spatial dimensions.

\section{Conclusions and perspectives}
\label{sec:conclusions}

In this work we provided a complete characterization of the dynamical scaling which emerges after a deep quench  of an \emph{isolated} quantum vector model with $O(N)$ symmetry at or below the point of its dynamical phase transition. The lack of intrinsic time and length scales is responsible for the occurrence of \emph{aging} phenomena similar to the ones observed in non-equilibrium classical systems~\cite{Janssen1988,Gambassi2005} or, more recently, in isolated~\cite{Chiocchetta2015} quantum many body systems.
While previous investigations of this phenomenon were based on a perturbative, dimensional expansion around the upper critical dimensionality $d_c=4$ of the model~\cite{Chiocchetta2015},
here we carry out our analysis within the exactly solvable (non-perturbative) limit $N\to \infty$, which allows us to obtain
exact results for scaling exponents and scaling functions of the relevant dynamical correlations, depending on the dimensionality $d$ of the model.
We find that the value of the pre-quench spatial correlation length (assumed to be small) controls the microscopic time $t_\Lambda$ after which the aging behavior emerges: in addition, it acts as an effective temperature for the dynamics after the quench, which, inter alia, determines a shift of the upper critical dimensionality $d_c$ of the model, as it occurs in equilibrium quantum systems at finite temperature~\cite{Sondhi97,Sachdevbook}.
Moreover, we provide evidence of the emergence of a  dynamic scaling behavior for quenches below $r_c$,  associated with coarsening, which we characterized numerically and analytically by studying the dependence of the relevant exponents on the spatial dimensionality $d$ of the system.

The exactly solvable model considered here provides a prototypical example of a dynamical phase transition (DPT) and of the associated aging occurring in a non-thermal stationary state. 
This state is expected to become unstable in 
systems with finite $N$, when the non-integrable terms of the Hamiltonian become relevant,
causing thermalization.
Nonetheless, this DPT might be still realized in the prethermal stage of the relaxation of actual quantum systems evolving in isolation from the surrounding environment~\cite{Berges2004,Kitagawa2011,Gring2012,Langen2013,Kollar2011,Moeckel2008,Moeckel2009,Moeckel2010,Marino2012,Marcuzzi2013,Mitra2013,VanDenWorm2013}. 
The latter are nowadays rather easily realized in trapped ultracold atoms, the behavior of which can be analyzed with remarkable spatial and temporal resolution~\cite{GreinerMandel02,Trotzky2008,Bakr2009,Bakr2010,Sherson2010,Endres2011,Cheneau2012,Fukuhara2013}.
In general, physical systems which can be described by some effective Hamiltonian with $O(N)$ symmetry include experimental realizations with ultra-cold atoms of the Bose-Hubbard model~\cite{Greiner2002,Bakr2009,Trotzky2008} (corresponding to $N = 2$) and one-dimensional tunnel-coupled condensates~\cite{Betz2011, Langen2013} ($N=1$).
In passing, we mention that an alternative and promising experimental realization of these models currently under investigation~\cite{Larre2014} is based on fluids of light propagating in non-linear optical media, which is expected to be ready for testing in the near future.

At least in principle, the universal dynamic scaling behavior emerging after a sudden quench which is highlighted in the present work can be experimentally studied by determining directly the two-time linear response and correlation functions of the system. Alternatively,  one can exploit the statistics of excitations produced after a (double) quench, as proposed in Ref.~\cite{Smacchia2014}.
In fact, the $n$-th cumulant $C_n(t)$ of the corresponding distribution was shown to grow as a function of the time $t$ elapsed since the quench, with a behavior which may saturate, grow logarithmically or algebraically, depending on $n$, $d$ and on whether the quench occurs above, at, or below criticality.
In particular, it was shown~\cite{Smacchia2014} that the increase in time of $C_n(t)$  is proportional to the integral over $\kk$ of the $n$-th power of the quantity $\rho_k(t)$ related to the number of excitations, which we briefly discussed  in the last paragraph of Sec.~\ref{sec:critical}. In turn, at long times, the leading growth of $\rho_k(t)$ is the same as the one of $|f_\kk(t)|^2$ (see the definition of $\rho_k(t)$ in Ref.~\cite{Smacchia2014}), i.e., of  $iG_K(k,t,t)$ (see Eq.~\eqref{eq:GK-f} here). As a result, a simple comparison with Eqs.~\eqref{eq:GK-short-time} and  \eqref{eq:GK-long-time} yields
\begin{equation}
C_n(t) \propto \int \!\!\dd^d k\, [iG_K(k,t,t)]^n \propto t^{n(1+2\alpha) - d},
\end{equation}
for a quench to the critical point; for a quench below it, instead, one finds the same expression with $\alpha$ replaced by $\alpha_\text{co}$, i.e.,
\begin{equation}
C_n(t) \propto \int \!\!\dd^d k \, [iG_K(k,t,t)]^n \propto t^{n(1+2\alpha_\text{co}) - d},
\end{equation}
where the values of the exponents $\alpha$ and $\alpha_\text{co}$ are given in Eqs.~\eqref{eq:a-analytic}, \eqref{eq:a-analytic-mf}, and \eqref{eq:exp-co}.
As a result, a measure of the statistics of the number of excitations for a quench would provide direct information on the aging and coarsening properties of the system.

The quantum aging and coarsening discussed in this work 
enrich the list of mechanisms underlying the 
scale-invariant non-thermal fixed points (NTFP)~\cite{Berges2008,Berges2009,Scheppach2010}, which have been so far interpreted in terms of quantum turbulence~\cite{Berges2011, Berges2012, PineiroOrioli2015} and dynamics of topological defects~\cite{Schole2012, Karl2013}. The extent to which 
these mechanisms are interconnected and combined in the dynamics of physical systems represents an intriguing yet challenging question for future investigations.


\acknowledgments
The authors thank I.~Carusotto, A.~Silva and P.~Smacchia for invaluable discussions. A. Mitra was supported by
National Science Foundation Grant No. NSF-DMR 1303177.



\appendix

\section{Comparison with aging after a classical quench}
\label{sec:comparison}

In this Appendix, we briefly review the phenomena of aging and coarsening in classical systems evolving in contact with a thermal bath after quenching its temperature either at or below a critical point of the system \cite{Cugliandolo1997}. 
In both cases, the lack of intrinsic time- and length-scales causes the emergence of algebraic behaviors in the temporal dependence of, e.g., two-time correlation and response functions which can be characterized in terms of scaling exponents and scaling functions with a certain degree of universality.  
The stochastic dynamics of these classical statistical systems can be simply described by effective models~\cite{Hohenberg1977}, which take the form of Langevin equations for the (coarse-grained) relevant degrees of freedom of the system. 
For example, a $N$-component real field $\vecvarphi = (\varphi_1,\dots,\varphi_N)$ obeying purely dissipative dynamics evolves according to the so-called model A
\begin{equation}
\label{eq:modelA}
\partial_t \varphi_a(\xx,t) = - D\frac{\delta {\mathcal H}[\vecvarphi]}{\delta \varphi_a(\xx,t) } + \zeta_a(\xx,t),
\end{equation}
where $D$ is a diffusion coefficient, 
$\mathcal{H}$ a $O(N)$-symmetric effective Hamiltonian in $d$ spatial dimensions
\begin{equation}
\mathcal H[\vecvarphi] = \int\dd^d x\, \left[\frac{1}{2}(\nabla\vecvarphi)^2 + \frac{r}{2}\vecvarphi^2 + \frac{u}{4! N}\vecvarphi^4 \right]
\label{eq:cl-H}
\end{equation}
and $\zeta_a$ is a zero-mean Gaussian white noise describing the thermal fluctuations of the reservoir, with correlations
\begin{equation}
\langle \zeta_a(\xx,t)\zeta_b(\xx',t') \rangle  = 2\,D\,T\, \delta_{ab}\,\delta^{(d)}(\xx-\xx')\,\delta(t-t').
\end{equation}
The evolution prescribed by Eq.~\eqref{eq:modelA} is such that the distribution of the fluctuating 
field $\vecvarphi$ at long times relaxes to the 
equilibrium distribution $P_\text{eq}[\vecvarphi] \propto \ee^{-\mathcal H[\vecvarphi]/T}$, independently  
of the initial condition. 
In this stationary state, and upon varying $r$, the system undergoes a second-order phase transition at 
the critical point $r=r_c$ of $\mathcal{H}$.  
Generically, $r$ in the effective Hamiltonian \eqref{eq:cl-H} is actually a function of the temperature $T$, with $r-r_c \propto T-T_c$, where $T_c$ is the critical temperature of the classical system.

A classical quench protocol~\cite{Janssen1988,Gambassi2005} consists in preparing the system in, e.g.,
a disordered state at high temperature (i.e., with vanishing correlation length) at $t=0$ and in letting it evolve with fixed $r =  r_c$ or $r < r_c$ for $t>0$. As a result, in both cases, the system relaxes to the equilibrium distribution with an algebraic behavior characterized by universal exponents which control, e.g., the scaling of the Fourier transform in space of the two-time and two-point correlation and response function $C(k,t,t')$ and $R(k,t,t')$, respectively. In particular, 
for a quench to the critical point, $C(k,t,t')$ and $R(k,t,t')$ can be calculated exactly in the limit $N\to \infty$~\cite{Janssen1988,Godreche2000} and they read (by rescaling time one can set $D=1$) for $2<d<4$:
\begin{equation}
\label{eq:response-critical}
R(k,t,t') = \theta(t-t') \left(\frac{t}{t'}\right)^{(4-d)/4} \ee^{-k^2 (t-t')},
\end{equation}
and 
\begin{equation}
\label{eq:corr-critical1}
C(k,t,t') = \frac{1}{k^2} F(k^2t, k^2 t'),
\end{equation}
with $F(x,y)$ a scaling function defined as $F(x,y) = (4xy)^{(4-d)/4}\int_0^{2\text{min}(x,y)} \dd t\, t^{(d-4)/2} \ee^{t-x-y}$.
Assuming for simplicity $t>t'$, the asymptotic expression of $C(k,t,t')$ at short times $t$, $t' \ll k^{-2}$ is
\begin{equation}
\label{eq:corr-critical2}
C(k,t,t') \simeq \frac{4}{d-2} t' \left(\frac{t}{t'}\right)^{(4-d)/4},
\end{equation}
while, at long times $t$, $t' \gg k^{-2}$, it reads
\begin{equation}
\label{eq:corr-critical3}
C(k,t,t') \simeq \left(\frac{t}{t'}\right)^{(4-d)/4} \frac{\ee^{- k^2(t-t')}}{k^2}.
\end{equation}
The effective classical dynamics prescribed by Eq.~\eqref{eq:modelA}  is actually relevant also for the quantum system with Hamiltonian \eqref{eq:Hamiltonian} investigated in the previous sections, when the field $\vecphi$ is linearly coupled to a bath of harmonic oscillators with ohmic spectral density and at equilibrium with temperature $T$. This \emph{open} quantum system is then characterized by an \emph{equilibrium} critical point at $r=r^\text{eq}_c(T)$ at which the microscopically large-distance, long-time properties are effectively described by the critical classical model described above~\cite{Kamenevbook2011}. 
In fact, a critical quench in this open quantum system was recently studied in Ref.~\cite{Gagel2014,Gagel2015} starting from a disordered pre-quench state and the scaling functions of $G_R$ and $G_K$ --- which correspond, respectively, to the classical $R$ and $C$ --- turn out to agree with Eqs.~\eqref{eq:response-critical},  \eqref{eq:corr-critical2}, and \eqref{eq:corr-critical3}.

The effect of having an isolated instead of an open quantum system is not only revealed by the different value of the dynamical exponent $z$ (1 and 2, respectively) but  also by the scaling form of the corresponding correlation and response functions at short times, where the actual value of $z$ does not appear explicitly. 
In fact, from Eqs.~\eqref{eq:GK-scaling}, \eqref{eq:a-analytic}, and \eqref{eq:exp-J} one finds that $i G_K (k=0,t,t') \sim (tt')^{d/4}$, which can be compared with  $C(k=0,t,t')$ in Eq.~\eqref{eq:corr-critical2}. While the dependence of the earliest time $t'$ is characterized by the same exponent $d/4$ this does not apply to the dependence on $t$ and in fact the overall scaling form is significantly different, as expected on the basis of the different scaling dimensions of the relevant fields~\cite{Janssen1988,Chiocchetta2015}.  
Comparing, instead, the corresponding short-time response functions in Eqs.~\eqref{eq:GR-short-time}  (see also Eq.~\eqref{eq:a-analytic}) and \eqref{eq:response-critical}, one finds that the dependence on the time $t'$ 
is characterized by opposite powers, whereas the ones on $t$ are seemingly unrelated. 
The classical model can be exactly solved in the limit $N\to \infty$ also if the quench occurs  from the disordered state to  below the critical point $r<r_c$~\cite{Godreche2000}; for $d>2$, the corresponding response function reads:
\begin{equation}
R(k,t,t') = \theta(t-t') \left( \frac{t}{t'} \right)^{d/4}\ee^{-k^2(t-t')} ,
\end{equation}
while the correlation function is (with $t>t'$)
\begin{equation}
\label{eq:corr-coarsening}
C(k,t,t') = M^2_\text{eq}\, (8\pi t')^{d/2}  \left( \frac{t}{t'} \right)^{d/4}\ee^{-k^2(t+t')},
\end{equation}
with $M_\text{eq}$ being the value that the order parameter would have in equilibrium in the system described by the post-quench classical Hamiltonian. Accordingly, this system exhibits coarsening in the strict sense \cite{Bray1994}, as 
Eq.~\eqref{eq:corr-coarsening} satisfies the dynamical scaling in the form indicated by Eq.~\eqref{eq:dynamical-scaling-hp} with the proper value $z_c=2$ of the coarsening exponent.

\bibliography{biblio}

\end{document}